\documentclass[sigconf]{acmart}

\usepackage{algorithm}
\usepackage[noend]{algpseudocode}

\usepackage{marginnote}

\usepackage{pgfplots}
\pgfplotsset{compat=1.18} 
\usepackage{pgfplotstable} 
\usepackage{subcaption} 
\usepackage{filecontents} 

\usepackage{textcomp} 

\usepackage{enumitem}

\usepackage{balance}
\usepackage{array}
\usepackage{makecell}

\AtBeginDocument{%
  }

\setcopyright{none}
\begin{document}

\title{TXSQL: Lock Optimizations Towards High Contented Workloads (Extended Version)}


\author{Donghui Wang}
\author{Yuxing Chen*}
\author{Chengyao Jiang}
\author{Anqun Pan}
\thanks{*Yuxing Chen is the corresponding author.}
\affiliation{%
  \institution{Tencent Inc.}
  \country{\{dorseywang,axingguchen,idavidjiang,}\\
  \country{aaronpan\}@tencent.com}
  }


\author{Wei Jiang}
\author{Songli Wang}
\author{Hailin Lei}
\author{Chong Zhu}
\author{Lixiong Zheng}
\affiliation{%
  \institution{Tencent Inc.}
    \country{\{vitalejiang,stanleewang,harlylei,}\\
  \country{teddyzhu,paterzheng\}@tencent.com}
  }

\author{Wei Lu}
\author{Yunpeng Chai}
\author{Feng Zhang}
\author{Xiaoyong Du}
\affiliation{%
  \institution{\mbox{Renmin University of China}}
  \country{\{lu-wei,ypchai,fengzhang,}\\
  \country{duyong\}@ruc.edu.cn}
}
\renewcommand{\shortauthors}{Wang et al.}

\begin{abstract}
%
Two-phase locking (2PL) is a fundamental and widely used concurrency control protocol. It regulates concurrent access to database data by following a specific sequence of acquiring and releasing locks during transaction execution, thereby ensuring transaction isolation. 
However, in strict 2PL, transactions must wait for conflicting transactions to commit and release their locks, which reduces concurrency and system throughput. We have observed this issue is exacerbated in high-contented workloads at Tencent, where lock contention can severely degrade system performance. While existing optimizations demonstrate some effectiveness in high-contention scenarios, their performance remains insufficient, as they suffer from lock contention and waiting in hotspot access.

This paper presents optimizations in lock management implemented in Tencent's database, TXSQL, with a particular focus on high-contention scenarios. First, we discuss our motivations and the journey toward general lock optimization, which includes lightweight lock management, a copy-free active transaction list, and queue locking mechanisms that effectively enhance concurrency. 
Second, we introduce a hotspot-aware approach that enables certain highly conflicting transactions to switch to a group locking method, which groups conflicting transactions at a specific hotspot, allowing them to execute serially in an uncommitted state within a conflict group without the need for locking, thereby reducing lock contention.
Our evaluation shows that under high-contented workloads, TXSQL achieves performance improvements of up to 6.5x and up to 22.3x compared to state-of-the-art methods and systems, respectively. 
\end{abstract}

\maketitle

\section{Introduction}\label{sec:introduction}
In Tencent Cloud \cite{tencent_cloud}'s Online Transaction Processing (OLTP) scenarios, databases typically experience low request loads most of the time. During these periods, the volume of user requests is below the system's processing capacity. 
However, there are occasional brief intervals characterized by sudden spikes in request traffic. 
For instance, in e-commerce, best-selling products or major promotional events can lead to surges in user visits and transaction volumes. These sudden spikes are often regarded as highly contented workloads, where the most frequently accessed data typically becomes a \textit{hotspot}. 
Although these workload transactions are typically short \cite{DBLP:conf/sosp/TuZKLM13,DBLP:journals/pvldb/KallmanKNPRZJMSZHA08-hstore,DBLP:conf/osdi/Lu00L23_NCC,DBLP:journals/pvldb/ZhangLL24_hybench,DBLP:journals/tkde/DongZLZ24_cloud_native,DBLP:journals/tkde/ZhangLZZF24_htap_survey}, the system's efforts to maintain transaction isolation result in numerous lock requests and contention, which negatively impacts performance \cite{DBLP:conf/sigmod/KossmannKL10,DBLP:journals/pvldb/AppuswamyAPIA17,DBLP:journals/pvldb/HuangQKLS20}. 


Pessimistic concurrency control protocols, such as two-phase locking (2PL) \cite{DBLP:journals/csur/BernsteinG81}, are generally considered more effective for managing high-conflict transactions \cite{DBLP:journals/pvldb/RenTA12_vll,DBLP:conf/eurosys/MuAS19_DRP}. However, traditional 2PL protocols often struggle to address scenarios involving hotspots (e.g., \cite{DBLP:journals/pvldb/BarthelsMTAH19,DBLP:journals/pvldb/HardingAPS17_deneva}). Consequently, several enhancement strategies have been proposed to mitigate these limitations.
An intuitive approach is to reduce the duration of lock holding. For instance, QURO \cite{DBLP:journals/pvldb/YanC16_QURO} introduces a commit-time-update mechanism that allows transactions to acquire locks later, thereby minimizing lock holding time. 
Escrow \cite{DBLP:journals/tods/ONeil86_escrow} is a locking mechanism that temporarily stores the access rights to resources through a third-party intermediary to reduce lock contentions.
Similarly, Bamboo \cite{DBLP:conf/sigmod/GuoWYY21_BAMBOO} proposes a violation of the 2PL protocol by releasing locks earlier to achieve the same goal. While these methods demonstrate some effectiveness in conflicted workloads, they cannot eliminate lock acquisitions and contention during hotspot access. In contrast, deterministic protocols (e.g., \cite{DBLP:conf/sigmod/ThomsonDWRSA12_calvin,DBLP:journals/pvldb/LuYCM20_aria,DBLP:conf/sosp/QinBG21_Caracal}) can effectively eliminate locking by scheduling transactions in a conflict-free manner. However, they face limitations due to scheduling overhead and the requirement for pre-acquisition of read-write sets, which are generally considered impractical \cite{DBLP:journals/jcst/DongTWWCZ20_DOCC}.

Most academic implementations of these novel protocols are found in in-memory database prototypes (e.g., DBX1000 \cite{DBLP:journals/pvldb/YuBPDS14_DBX1000}), while industry practitioners tend to be cautious about modifying concurrency control protocols. For example, systems like MySQL and SQL Server rely on traditional 2PL and multi-version concurrency control (MVCC) \cite{DBLP:journals/tods/BernsteinG83}, or a combination of both. Also, rather than altering the transaction layer to mitigate hotspot workloads, they often focus on modifying the application layer. 
For instance, most of our Tencent Cloud database instances implement request restrictions at the application layer to prevent sudden spikes in hotspot requests (for further details, see Section \ref{sec:implementation_fix_reqesut_latency_optimization}). Additionally, some databases employ proactive hotspot identification and SQL rewriting \cite{DBLP:journals/pvldb/GuravannavarS08_query_rewrite} at the application layer to address ad-hoc high-contented workloads (e.g., \cite{polardb_website}). However, these implementations may not necessarily represent the most effective solutions for managing hotspot access in commercial databases, and in certain dynamic scenarios, they may not be applicable at all.

In certain high-contented workloads at Tencent, we have found that request restrictions can effectively mitigate excessive lock contention. However, this approach is less effective in dynamic hotspot situations (see Section \ref{sec:typical_application}), where transactions concentrate on updating one or a few data items. In these cases, executing transactions serially may yield better performance by eliminating lock contention and associated overhead. To offer a practical solution for managing hotspots, this paper provides insights into lock optimization within Tencent Database TXSQL \cite{txsql}.
First, we discuss our motivations and the overarching lock optimization process, including lightweight lock management, a lock-free active transaction list, and a queue lock mechanism. These optimizations enhance concurrency and improve performance in typical high-contention scenarios. Secondly, we place particular emphasis on its optimizations for hotspot situations. Specifically, TXSQL is designed to meet the following two criteria for concurrency control when addressing hotspots: 
(1) the protocol is workload-agnostic and does not interfere with existing application logic; and (2) the protocol maintains performance when concurrency increases.


Inspired by the benefits of deterministic protocols in scheduling high-conflicted transactions, we propose a group locking mechanism within the 2PL protocol framework for managing hotspots. Unlike traditional 2PL, which treats all data uniformly, our approach distinguishes between hotspot and non-hotspot data. We group transactions updating hotspot data, allowing them to proceed without locking, as long as they adhere to a total order. 
When there are no hotspot updates, TXSQL reverts to traditional 2PL, ensuring minimal impact on overall throughput. 
Since 2023, we have successfully upgraded over 20,000 database instances of financial applications from MySQL to TXSQL, achieving over 30\% performance improvements in non-hotspot scenarios. In hotspot scenarios, the upgrade has not only reduced jitter effectively but also resulted in performance improvements of up to 10x.
Our main contributions are as follows:
\begin{itemize}
    \item We provide motivation, optimizations, and insights in TXSQL to address lock conflict issues.
    \item When involving hotspot data, we propose a group locking mechanism that groups hotspot data accesses and executes them serially without locking. Also, we describe its correctness in terms of deadlock, rollback, and failure recovery.
    \item Compared to state-of-the-art methods and systems, evaluation shows that our approach achieves performance improvements of up to 6.5x and 22.3x, respectively. 
\end{itemize}

\section{Preliminary}\label{sec:preliminary}
This section introduces the basic 2PL design, as well as TXSQL's system architecture and its applications.

\subsection{Two-Phase Locking (2PL)}
The 2PL protocol, widely implemented in many database management systems, is designed to prevent data conflicts between concurrent transactions. Transaction execution is divided into two distinct phases: the growing phase and the shrinking phase. In the growing phase, a transaction can acquire locks and access data but is prohibited from releasing any locks. In the shrinking phase, a transaction may release locks but cannot acquire any new locks. The 2PL ensures transactions do not conflict with one another during execution, thereby maintaining transaction isolation. 

Mutual Exclusion (Mutex) is a synchronization mechanism used in multithreaded programming, designed to prevent multiple threads from simultaneously accessing shared resources, thereby avoiding data races and inconsistencies. A mutex ensures that only one thread can access a specific resource or code segment at any given time. It is commonly utilized in 2PL lock management. 

\subsection{Transaction Execution Workflow}
TXSQL \cite{txsql_repo}, an open-source MySQL branch maintained by Tencent Cloud, is fully compatible with MySQL's syntax and APIs. 
\begin{figure}[t]
  \centering
  \includegraphics[width=1\linewidth]{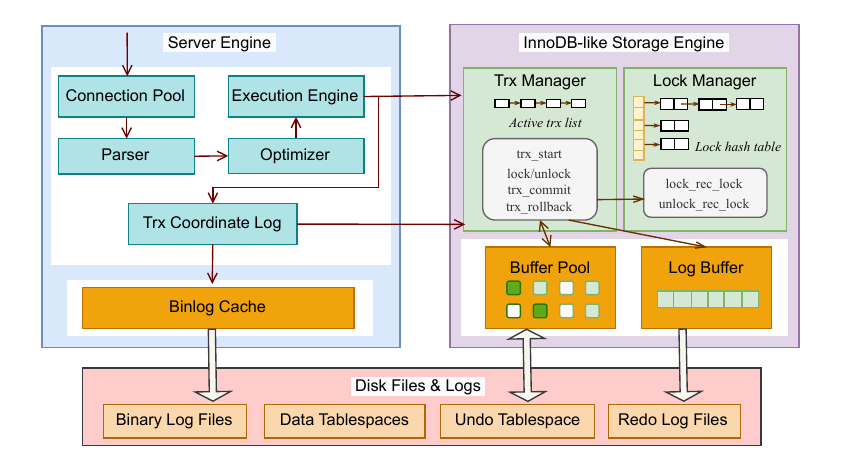}
  \caption{The core architecture of TXSQL.}
  \label{figue:system_architecture}
\end{figure}
Figure \ref{figue:system_architecture} depicts the core system architecture of TXSQL. The primary optimizations discussed in this paper focus on the transaction manager and the lock manager at the Storage layer. The transaction manager is responsible for the lifecycle of transactions, including their initiation, execution, commit, and rollback. It ensures the atomicity of transactions, meaning that they either complete successfully in their entirety or do not execute at all. In contrast, the lock manager implements a locking mechanism to regulate access to database resources during transaction execution, such as rows and tables. 

When a transaction initiates a row update in the transaction manager, the specific row can be uniquely identified by locating the record's associated tablespace through the $space\_id$, identifying the page containing the record via the $page\_no$, and pinpointing the exact position of the record within a page using the $heap\_no$. Therefore, the combination of <$space\_id$, $page\_no$, $heap\_no$> serves as a unique identifier for a row.
Once the corresponding record is located, the system first checks the lock manager whether any active transactions conflict with the lock that the current transaction intends to acquire. Locks created by all active transactions are managed by a hash lock table, where the key is generated from the record's <$space\_id$, $page\_no$>, and the value is the lock object ($lock\_t *$). If no conflict exists, the lock is successfully acquired, its status is set to SUCCESS, and it is inserted into the hash table $lock\_sys$. If a conflict is detected, the transaction is placed in a waiting state and added to a wait queue in $lock\_sys$. Once the transaction holding the lock commits and releases the lock, it will awaken the waiting transactions in $lock\_sys$, allowing the awakened transaction to acquire the lock. This process enables multiple transactions to operate simultaneously without interference.

\subsection{Typical Applications at Tencent}\label{sec:typical_application}
\noindent \textbf{High contented vs. Hotspot.}
High-contented workloads occur in concurrent environments where multiple transactions compete for access to the same resources, such as locks, data items, or database rows, leading to transaction delays. A hotspot refers to specific data items that are accessed frequently, causing them to become ``hot''. Hotspots typically focus on a limited number of items, representing an extreme form of high-contented workloads.

\noindent \textbf{WeChat Red Envelope} is a popular application on WeChat, which has a monthly active user base exceeding 1.3 billion. TXSQL has facilitated WeChat's red envelope payment system, allowing users to send red envelopes (monetary gifts) to others via WeChat. On traditional Chinese New Year's Eve, we successfully handled a peak of Transaction per Second (TPS) up to 14 million red envelopes, showing the system's capability to manage high volumes of concurrent transactions effectively.

\noindent \textbf{Tencent Financial Technology (FiT)} serves as a financial platform, offering a range of mobile payment and financial services. 
Its primary application is WeChat Pay, which facilitates an average of over one billion payment transactions daily. These transactions encompass a wide array of daily activities, including shopping, dining, transfers, and entertainment. During holidays or special events, such as Double Eleven Day, transaction volumes on WeChat Pay typically experience a significant surge.

\noindent 
\textcolor{black}{\textbf{E-commerce applications} continue to exhibit hotspot scenarios, which are increasingly becoming the norm due to the business model of live-streaming. 
The number of active e-commerce influencers has reached 3 million \cite{Dong2024-live-stream-users}. Influencers typically offer products that feature significant price advantages but limited stock. These high-demand products attract a large number of customers who rush to purchase them, thereby making product inventory a focal point of interest. }

\section{Motivation and Design}\label{sec:design}
This section begins by presenting some general lock optimization techniques in TXSQL, followed by a discussion of the design and motivation underlying both initial and current hotspot lock mechanisms within TXSQL.
\subsection{General Lock Optimizations}\label{sec:design_general_lock_optimization}


\subsubsection{Lightweight Locking}
In the vanilla InnoDB Storage engine, table locks and row locks are the primary lock types, with all concurrent transaction lock information managed by a global lock system. 
This lock system has two significant shortcomings:
(1) It created a large number of row locks. For instance, 100 concurrent transactions, each updating 10 rows, create up to 1,000 lock records for low- or non-contented workloads.
(2) In high-contented workloads, even with partitioning optimizations \cite{DBLP:conf/edbt/JohnsonPHAF09_ShoreMT,DBLP:journals/pvldb/ChenPLYHTLCZD24_TDSQL} applied, acquiring the sharded mutex for the corresponding page still incurs a substantial amount of lock wait time (see Figure \ref{fig:effect_optimizations_lock_wait}). 

The goal of lightweight lock optimization is to streamline lock logic by minimizing the number of locks created and lowering the overhead associated with maintaining row locks. To achieve this, we introduced a variable named $trx\_lock\_wait$ to manage transactions waiting for locks. This variable utlizes a lock-free hash design implemented with a map container, where the record ID serves as the key and the record value is a queue of waiting transactions.  By recording transaction IDs, we can swiftly identify conflicts, creating locks only when necessary, thereby effectively reducing lock overhead. 

\subsubsection{Copy-Free Active Transaction List}
During concurrent transaction execution, it checks the transaction IDs of all active transactions to determine which version of the data should be read in MVCC\footnote{\textit{readView} structure is employed in InnoDB for MVCC.}. Directly accessing the active transaction list can lead to performance issues, as it necessitates locking the list each time data is read.
To mitigate this problem, one intuitive solution is to create a copy of the active transaction list during data reads, allowing checks on the copied list without locking the original. However, when read-write conflicts are substantial, the overhead associated with maintaining these copies can still be considerable.

To address this, we leverage snapshots in MVCC to avoid direct copying and locking of the active transaction list.
Specifically, we aim to determine which data is visible by comparing the transaction ID with the version information of the data. We introduce a new attribute, $del\_ts$, for each transaction ID, which indicates the deletion timestamp of each transaction. This attribute aids in assessing the transaction's status, and now visibility can be determined by snapshot information and the $del\_ts$ attribute without acquiring the active transaction list, thus improving concurrency. Detailed implementation can be found in \cite{txsql_repo}.

\pgfplotstableread{
1	11200.43	12150.76	11712.38	23892.52
2	9651.97	11230.78	10076.62	26730.67
3	5994.18	9513.44	9928.07	24438.96
4	2627.1	6523.3	8004.52	19958.37
}\sysbenchhotspot

\pgfplotstableread{
1	52495.38	56032.77	54148.01	54011.05
2	99696.27	104464.68	104040.39	103577.95
3	104446.66	107730.14	107067.34	107228.84
4	85273.93	89967.6	89115.43	89658.34
}\sysbenchuniform

\pgfplotstableread{
1	6415.02	6520.9	6498.61	6525.57
2	5837.39	6186.73	6243.99	6233.98
3	2815.77	5232.24	5173.5	5159.15
4	839.56	4110.87	4119.28	4103.23
}\sysbenchscan

\pgfplotstableread{
1	6128.97	7133.67	7079.8	7173.28
2	21112.83	24577.2	24240.45	24576.71
3	33348.2	38052.15	37970.59	37954.41
4	33403.29	38110.6	38183.54	38135.22
}\sysbenchreadonly

\pgfplotstableread{
1	71040.3	80914.16	58806.33	96114.07
2	48460.81	79378.63	52177.4	94736.59
3	29389.34	73569.1	43044.92	88605.37
4	23107.62	68982.99	39582.9	83584.65
5	19433.6	62500.21	37140.45	77286.64
}\sysbenchskewness

\pgfplotstableread{
1	4220.85	9958.35	13100.09	25480.61
2	3819.13	5710.66	6871.02	21537.64
3	2935.24	3713.37	4211.9	15414.67
4	1781.33	2141.86	2217.23	10448.31
}\sysbenchlength


\pgfplotstableread{
1	29606.12	32009.1	31730.95	31982.82
2	2330.12	2773.54	3130.18	11480.9
3	1583.86	1771.02	1921.6	7944.65
4	1286.77	1302.82	1407.71	4483.47
}\sysbenchreadwriteratio

\pgfplotstableread{
1	11200.43	6412.3	16623.17	23892.52
2	10446.67	6544.18	16140.55	28126.87
3	9651.97	6644.54	14180.44	26730.67
4	8894.45	6725.12	12579.41	26180.5
5	7756.42	6929.41	10357.31	25711.16
6	5994.18	7203.62	7619.68	24438.96
7	4149.99	7318.7	5008.5	22847.13
8	2627.1	7851.32	3058.34	19958.37
}\sysbenchscalability
\pgfplotstableread{
1	0.75	10.27	0.55	0.35
2	1.61	20.37	1.06	0.57
3	3.49	20.37	2.39	1.12
4	7.43	40.37	5.37	2.22
5	17.01	70.55	13.22	4.49
6	44.17	130.13	37.56	9.39
7	130.13	223.34	110.66	20.74
8	411.96	419.45	356.7	125.52
}\sysbenchscalabilitylatency

\pgfplotstableread{
1	6796.73
2	11382.42
3	12338.51
4	11200.43
5	10446.67
6	9651.97
7	8894.45
8	7756.42
9	5994.18
10	4149.99
11	2627.1
}\sysbenchupdatemotivation

\pgfplotstableread{
1	4149.99	7867.54  22459.88
2	2752.68	3730.41  18322
3	1719.96	2027.45  16462.34
4	1013.06	1114      13493.87
5	552.03	584        5097.02
}\sysbenchupdatelengthmotivation

\pgfplotstableread{
2	-0.01	-0.01	0.0705 	0.0598 
3	-0.01	-0.01	0.1520 	0.1190 
4	-0.01	-0.01	0.3672 	0.2250 
5	-0.01	-0.01	0.4773 	0.3409 
}\sysbenchabortrate

\pgfplotstableread{
1	6436.67	1117.23	2758.63
2	15105.68	2598.91	6623.16
3	17229.89	4517.76	10906.42
4	17147.56	1981.13	12075.4
5	17081.68	1592.08	11990.73
}\sysbenchbatchsizeone

\pgfplotstableread{
1	8263.56	1249.84	3492.33
2	18882.51	2754.36	7492.66
3	20879.03	5054.38	10296.73
4	20654.51	1958.84	10273.41
5	20879.03	1958.84	5082.71
}\sysbenchbatchsizetwo

\begin{figure}
    \centering
    \begin{subfigure}[b]{0.235\textwidth}
    \begin{tikzpicture}
        \begin{axis}[
            width=4cm,
            height=3cm,
            xlabel={Thread},
            ylabel={TPS(Txn/s)},
            xtick={1,4,7,11},
            xticklabels={1,8,64,1024},
            ymin=0, 
            legend style={at={(1,1.05)}, anchor=south east},
            grid=major,
            bar width=0.5,
            enlarge x limits=0.125,
            ybar=2*\pgflinewidth,
            every ybar/.style={fill=blue!30},
            legend columns=4,
            y tick label style={scaled ticks=base 10:0},
            tick align = outside,
        ]

        \foreach \i in {1} {
            \addplot table[x index=0, y index=\i] {\sysbenchupdatemotivation};
        }
        \end{axis}
    \end{tikzpicture}
    \caption{SysBench hotspot update}
    \label{fig:design_motivation_mysql}
    \end{subfigure}
    \begin{subfigure}[b]{0.235\textwidth}
    \begin{tikzpicture}
        \begin{axis}[
            width=4cm,
            height=3cm,
            xlabel={Transaction length},
            ylabel={TPS(Txn/s)},
            xtick={1,2,3,4,5},
            xticklabels={1,2,4,8,16},
            ymin=0, 
            legend style={at={(1,1.05)}, anchor=south east},
            grid=major,
            bar width=0.15,
            enlarge x limits=0.125,
            ybar=2*\pgflinewidth,
            every ybar/.style={fill=blue!30},
            legend columns=4,
            y tick label style={scaled ticks=base 10:0},
            tick align = outside,
        ]

        \foreach \i in {1,2,3} {
            \addplot table[x index=0, y index=\i] {\sysbenchupdatelengthmotivation};
        }
        \legend{MySQL,Queue, Group}
        \end{axis}
    \end{tikzpicture}
    \caption{SysBench hotspot update}
    \label{fig:design_motivation_lock}
    \end{subfigure}
    \caption{Motivation of hotspot optimization.}
    \label{fig:design_motivation}
\end{figure}
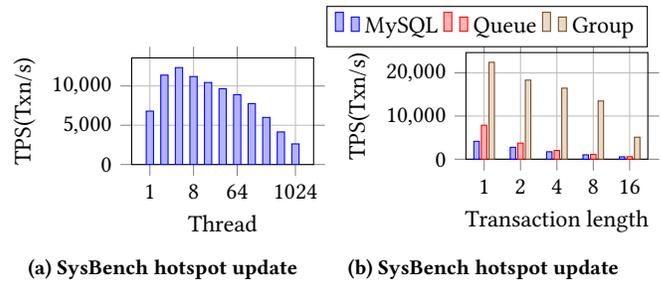

\subsection{Quene Locking for Hotspot Access}\label{sec:design_queue_locking} 
We have identified performance issues associated with updates to hotspot rows, particularly when multiple transactions attempt to update the same row simultaneously.
Notably, testing MySQL with a SysBench hotspot update workload (setup in Section \ref{sec:evaluation_setup}) at a concurrency level of 1024 showed worse performance than serial execution at a concurrency level of 1, as illustrated in Figure \ref{fig:design_motivation_mysql}.
Analysis revealed that this issue stems from the lock wait queue in the lock hash table $lock\_sys$ for each data row, where the cost of deadlock detection increases with the length of the queue. Since deadlock detection is invoked during each transaction, and locks $lock\_sys$, blocking other transactions. Consequently, as concurrency increases, the length of the lock wait queue also grows, leading to higher deadlock detection costs. 

A common approach is to reduce the locks holding time of transactions (e.g., QURO~\cite{DBLP:journals/pvldb/YanC16_QURO}), which often (a) employs commit-time updates to defer write lock occupancy or (b) utilizes timestamp splitting to separate read and write timestamps.
Another approach involves implementing a lightweight strategy to manually restrict access to hotspot data rows, similar to PolarDB's solution \cite{polardb_website} by adding a hint to transaction hotspot query. However, these approaches necessitate modifications to the transaction syntax. For example, QURO requires reordering queries, while PolarDB adds hints for hot queries, meaning 
they cannot automatically detect hotspot updates during runtime for dynamic workloads.

We propose a queue locking mechanism specifically designed for hotspots. To enhance its generality and practicality, our design avoids syntax modifications and can automatically detect hotspot updates by monitoring row update operations and collecting update information in a lightweight manner (details in Section \ref{sec:implmentation_detect_hotspots}).
Once a hotspot row is detected, its unique identifier is inserted as a key into a hotspot hash table, with the corresponding value being a queue that holds waiting transactions. Subsequently, when a transaction requests a row, if the unique identifier of that row is present in the hotspot hash table, the transaction is queued and will be awakened by preceding transactions that release their locks upon committing. Unlike traditional locking mechanisms, transactions updating hotspot rows do not directly acquire locks from the lock manager. Instead, they queue up before attempting to lock, thereby avoiding the overhead associated with contention for the lock manager.

To mitigate potential deadlocks during hotspot updates, we implemented a timeout mechanism instead of direct deadlock detection for two main reasons: (1) Performance: Deadlock detection requires traversing all related transaction information, which our tests (see Figure \ref{fig:design_motivation_mysql}) indicated is less efficient in hotspot scenarios. The timeout mechanism generally performs better and allows for configurable transaction wait times, making it more adaptable to dynamic real-world conditions.  
(2) Code complexity: Integrating deadlock detection would increase code complexity, complicating maintenance and raising the risk of new errors and instabilities. 


\subsection{Group Locking for Hotspot Access}\label{sec:design_group_locking}
We achieved improved results with the queue locking mechanism, however, its effectiveness was limited in customer workloads due to full synchronization between primary and secondary servers required for high availability, which increased transaction latency. Our analysis, using the SysBench workload with varying transaction lengths, confirmed that longer transaction latencies diminish queue locking effectiveness, as shown in Figure \ref{fig:design_motivation_lock}.

We further survey current approaches and categorize them into three types of scheduling:
(1) Thread-level scheduling~(e.g., \cite{DBLP:journals/pvldb/RenTA12_vll}): This approach schedules thread utilization through a thread pool to avoid thread-level contention.
(2) Transaction-level scheduling (e.g., 
\cite{DBLP:conf/vldb/KimuraGK12_early_lock_release,DBLP:conf/sigmod/JungHFHY13,DBLP:conf/sigmod/GraefeLKTV13_CLV,DBLP:journals/pvldb/YanC16_QURO,DBLP:conf/eurosys/ChenSJRLWZCC21_DAST}): Similar to queue locking in Section \ref{sec:design_queue_locking}, this method minimizes lock holding time to mitigate lock contention.
(3) Query-level scheduling (e.g., \cite{DBLP:conf/sosp/ZhangPZSAL13_chopping,DBLP:conf/sigmod/FaleiroTA14_lazy,DBLP:conf/sosp/XieSLAK015_Callas,DBLP:journals/pvldb/DingKG18,DBLP:journals/pvldb/LuYCM20_aria}): This includes deterministic protocols that eliminate lock contention and transaction chopping that enables early reads of dirty data. 
For (1), they are orthogonal to our optimization focus in this paper.
For (2), their benefits diminish significantly in high-latency transactions as investigated in Figure \ref{fig:design_motivation_lock}, since locks can only be obtained after the conflicting transaction commits. 
For (3), they are generally impractical for conventional cloud service providers like us, as pre-obtaining user read-write sets is typically challenging. Moreover, in non-conflict situations, their performance often lags behind non-deterministic protocols due to scheduling overhead \cite{DBLP:journals/pvldb/HardingAPS17_deneva}.

\begin{figure}[t]
  \centering
  \begin{subfigure}[t]{0.15\textwidth}
  \includegraphics[width=1\linewidth]{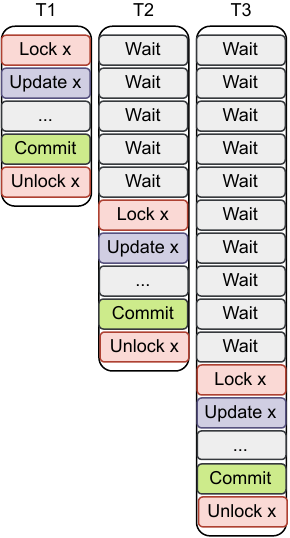}
  \caption{2PL}
  \label{figue:design_group_lock_2pl}
  \end{subfigure}
  \begin{subfigure}[t]{0.15\textwidth}
  \includegraphics[width=1\linewidth]{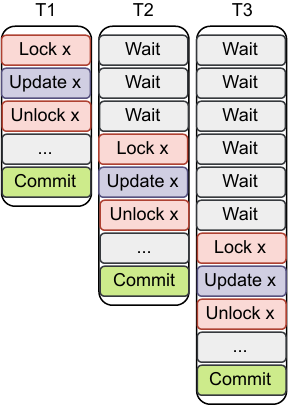}
  \caption{Bamboo}
  \label{figue:design_group_lock_bamboo}
  \end{subfigure}
  \begin{subfigure}[t]{0.15\textwidth}
  \includegraphics[width=1\linewidth]{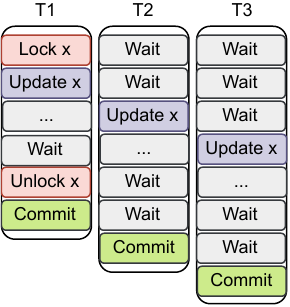}
  \caption{Group locking}
  \label{figue:design_group_lock_group_lock}
  \end{subfigure}
  \caption{Locking schedules of update a hotspot row x under (a) 2PL,  (b) Bamboo \cite{DBLP:conf/sigmod/GuoWYY21_BAMBOO}, and (c) group locking.}
  \label{figue:design_group_lock_comparison}
\end{figure}

The state-of-the-art hotspot-oriented protocol that improves 2PL is Bamboo \cite{DBLP:conf/sigmod/GuoWYY21_BAMBOO}, which violates 2PL by releasing locks before commit as depicted in Figure \ref{figue:design_group_lock_bamboo}. In contrast, traditional 2PL requires locks to be released only after commit, as depicted in Figure \ref{figue:design_group_lock_2pl}. While Bamboo is effective for certain high-contented workloads, it still incurs overhead from acquiring and releasing locks for each update
To address this limitation, we, inspired by deterministic protocols, propose a group locking mechanism. This mechanism logically groups a set of updates on a conflicting hotspot data row, allowing these updates to be executed sequentially within the group without locking. As depicted in Figure \ref{figue:design_group_lock_group_lock}, locking and unlocking operations occur only once per group, enabling conflicting transactions to proceed without locks. This approach eliminates the need for lock release and wake-up processes between transactions within the group, significantly reducing the time spent waiting for locks. 

Note that all current solutions and our solution only update a single hotspot row per transaction and lack support for transactions with multiple hotspot rows. Such concurrent multi-row hotspot transactions often trigger deadlocks, resulting in cascading rollbacks and poor performance. In practice, transactions usually update one hotspot at a time. 
\section{Implementation}\label{sec:implementation}
This section outlines the implementation of group locking in TXSQL and addresses various challenges and optimizations.
Section \ref{sec:implmentation_detect_hotspots} focuses on the detection and representation of hotspot rows, while Section \ref{sec:implementation_transaction_processing} details the transaction processing mechanisms for group locking. Ensuring correctness in transaction processing between hotspot and non-hotspot data rows is crucial, and this is addressed in Sections \ref{sec:implementation_commit_order} and \ref{sec:implementation_rollback_order}, which detail how TXSQL maintains correct commit and rollback orders. Section \ref{sec:implementation_deadlock_detection} discusses the deadlock prevention mechanism, and Section \ref{sec:implementation_fix_reqesut_latency_optimization} presents latency optimizations to tackle challenges arising from group locking.

\subsection{Management of Hotspots}\label{sec:implmentation_detect_hotspots}
Recall the locking process: if a transaction attempts to acquire a lock that conflicts with an active transaction, it will be added to the corresponding lock wait queue. A longer wait queue for a specific row indicates higher contention, suggesting that the row is a hotspot. 
We define a row as a \textit{\textbf{hotspot}} when the number of waiting transactions exceeds a threshold (e.g., 32, which is a rule of thumb \cite{DBLP:journals/pvldb/TianHMS18}).
Once identified, the hotspot is added to a hotspot hash table, denoted as $hot\_row\_hash$, and subsequent update transactions for this row must wait in the queue.
To manage these entries effectively, a background thread periodically monitors the $hot\_row\_hash$. 
If an entry is no longer a hotspot (e.g., no waiting transactions), it is removed from the $hot\_row\_hash$, and future updates revert to the standard 2PL protocol. This simple yet effective mechanism enables us to identify and manage hotspot data with minimal overhead.

\begin{figure}[t]
  \centering
  \includegraphics[width=1\linewidth]{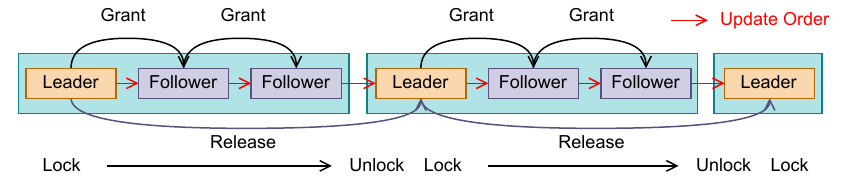}
  \caption{Design of group locking.}
  \label{figue:design_group_lock}
\end{figure}

\subsection{Transaction Processing}\label{sec:implementation_transaction_processing}
The concurrently arriving transactions on a hotspot row are automatically organized into groups, as depicted in Figure \ref{figue:design_group_lock},. The key design of the group lock is as follows:
\begin{itemize}[left=0em]
    \item \textbf{Transaction role:} Within a group, transactions are either a leader or followers. The initial transaction serves as the leader, responsible for lock acquisition and release. The remaining transactions as followers do not participate in any locking operations.
    \item \textbf{Transaction execution:} Within a group, hotspot updates execute serially. Once a hotspot update is completed, it automatically grants execution to the next follower. Queries except those updating hotspot rows are executed in parallel.
    \item \textbf{Lock management:} When the leader transaction commits, it releases the row lock and awakens the next leader. The current leader releases the lock only after the last update within the group is completed. Subsequently, the new leader will acquire the row lock, thereby forming a new group.
\end{itemize}

During execution, \textit{\textbf{a dependency list}} is formed and maintained based on the update order, which determines the order of commits and rollbacks. As queries are executed, each hotspot update transaction is assigned a globally incrementing identifier, referred to as $hot\_update\_order$, which is added to the dependency list, establishing a sequence of dependencies. A transaction can commit (or rollback) if it has no preceding (or subsequent) transaction in the dependency list. More details on commit and rollback order will be discussed in Sections \ref{sec:implementation_commit_order} and \ref{sec:implementation_rollback_order}, respectively.

The transaction execution process, as shown in Algorithm \ref{alg:transaction_execution}, involves acquiring locks and queuing transactions, updating, and awakening transactions, as follows:

Step 1 - Locking and queuing (Lines 2-9): Check whether the item is a hotspot and whether it has a dependency list. If both conditions are met, the transaction is added to the waiting queue for updates and will remain in a waiting state until awakened. Otherwise, the item is locked by 2PL locking. If the transaction's hot update status is not empty, it is added to the item's dependency list, and the global hot update order is updated.

Step 2 - Updating (Lines 10-14): Update the item and the state of the transaction. If the transaction's state is GRANTED, the item will no longer grant access to a new follower transaction. If the transaction's hot update status is RUNNING and the transaction is the leader, the item will not switch to a new leader.

Step 3 - Awakening the next leader/follower (Lines 15-20): If the item is switching to a new leader, the process returns. Otherwise, the next transaction is retrieved from the waiting queue, the item's state for granting access to the new transaction is set to true, the hot update status of the next transaction is marked as GRANTED, and the next transaction is awakened.

\begin{algorithm}[t]
\caption{Transaction Execution Process}
\label{alg:transaction_execution}
\begin{algorithmic}[1]
\small
\Function{Execute}{T, item}
    \If{item is in hot\_spots \textbf{and} item.dep\_list $\neq$ 0}
        \State item.waiting\_updates.push\_back(T)
        \State os\_event\_wait(T.event) \Comment{\textcolor{gray!80}{Wait to be awakened}}
    \Else
        \State lock\_rec\_add(item) \Comment{\textcolor{gray!80}{Traditional 2PL locking}}
    \EndIf
    \If{T.hot\_update\_state $\neq$ NONE}
        \State trx.hot\_update\_order $\gets$ global\_hot\_update\_order.fetch\_add(1)
        \State item.dep\_list.push\_back(T) \Comment{\textcolor{gray!80}{Update dependency list}}
    \EndIf

    \State update(T, item) \Comment{\textcolor{gray!80}{Execute update}}
    \If{T.status = GRANTED}
        \State item.granting\_new\_trx $\gets$ false
    \EndIf
    \If{T.hot\_update\_status = RUNNING \textbf{and} T.is\_leader}
        \State item.switching\_new\_leader $\gets$ false
    \EndIf

    \If{item.switching\_new\_leader} 
        \State \Return  \Comment{\textcolor{gray!80}{To next leader}}
    \EndIf
    \State next\_trx $\gets$ item.waiting\_updates.front()
    \State item.granting\_new\_trx $\gets$ true
    \State next\_trx.hot\_update\_status $\gets$ GRANTED
    \State os\_event\_set(next\_trx.event)
\EndFunction
\end{algorithmic}
\end{algorithm}

\noindent \textbf{Switching to hotspot.}  
Upon initially transitioning to group locking, the older waiting transactions remain in the waiting queue of the $lock\_sys$. Once a transaction is identified as a hotspot row, any new conflicting transactions are queued in the $waiting\_update$ list of $hot\_lock\_sys$. The leader of the group locking mechanism prioritizes awakening the older waiting transactions from $lock\_sys$. Only after all these transactions have been processed does the leader begin to awaken the new waiting transactions from $waiting\_update$.

\subsection{Commit Order Guarantee}\label{sec:implementation_commit_order}
In group locking, we ensure that the commit order aligns with the update order through the use of the maintained dependency list.
The transaction commit process, detailed in Algorithm \ref{alg:transaction_commit}, consists of several key steps: releasing locks, awakening the next leader, and committing the transaction, as follows:

Step 1 - Release locks and awaken the next leader (Lines 2-10): Set the switching state of the new leader for the item to \textit{true}. Then, wait until the last granted transaction is completed. 
Next, iterate through all locks associated with the transaction and process each lock accordingly. 
If the lock can be released successfully, then continue. Next, retrieve the next transaction from the waiting updates queue, set its hot update status to RUNNING, and wake it up. Simultaneously, designate this next transaction as the leader. 

Step 2 (Lines 11-12) - Commit transaction: Commit the transaction and remove it from the dependency list of the item.

\begin{algorithm}[t]
\caption{Transaction Commit Process}\label{sec:implementation_transaction_commit}
\label{alg:transaction_commit}
\small
\begin{algorithmic}[1]
\Function{Commit}{T, item}
    \State item.switching\_new\_leader $\gets$ true \Comment{\textcolor{gray!80}{No more granting new trx}}
    
    \While{item \textbf{and} item.granting\_new\_trx}
        \State ut\_delay(10)  \Comment{\textcolor{gray!80}{Wait updates complete for all granted trx}}
    \EndWhile
    
    \For{lock \textbf{in} T.locks} 
        
        \If{lock\_rec\_release(lock)}
            continue
        \EndIf
        
        \State next\_trx $\gets$ item.waiting\_updates.front() \Comment{\textcolor{gray!80}{Get next trx}}
        \State next\_trx.hot\_update\_status $\gets$ RUNNING 
        \State os\_event\_set(next\_trx.event)
        \State next\_trx.is\_leader $\gets$ true
    \EndFor
    
    \State commit\_trx(T)
    \State item.dep\_list.erase(T) \Comment{\textcolor{gray!80}{Remove from dependency list}}
\EndFunction
\end{algorithmic}
\end{algorithm}

\begin{figure}[t]
  \centering
  \begin{subfigure}[t]{0.15\textwidth}
  \includegraphics[width=1\linewidth]{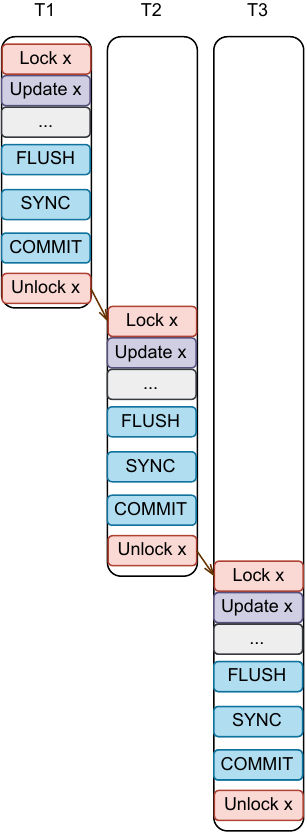}
  \caption{MySQL 2PC}
  \label{figue:design_group_lock_2pc_mysql}
  \end{subfigure}
  \begin{subfigure}[t]{0.15\textwidth}
  \includegraphics[width=1\linewidth]{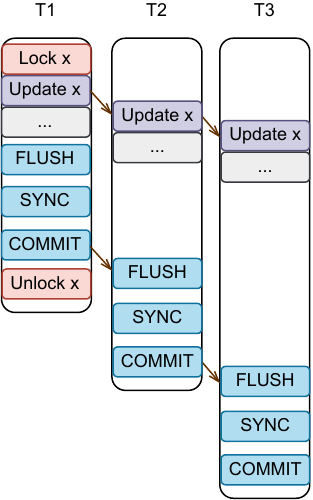}
  \caption{No group~commit}
  \label{figue:design_group_lock_2pc_wo}
  \end{subfigure}
  \begin{subfigure}[t]{0.15\textwidth}
  \includegraphics[width=1\linewidth]{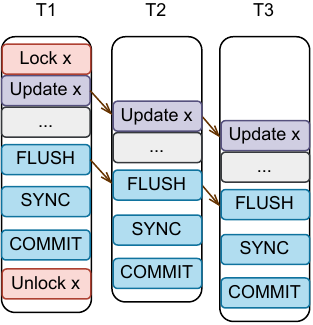}
  \caption{Group~commit}
  \label{figue:design_group_lock_2pc_w}
  \end{subfigure}
  \caption{Locking schedules of commit phase in 2PC for updating a hotspot row x under (a) MySQL, (b) TXSQL W/O group commit, and (c) TXSQL with group commit.}
  \label{figue:design_group_lock_2pc}
\end{figure}

\noindent \textbf{Group commit optimization.} 
In TXSQL, the XA/2-phase commit (2PC) mechanism is employed to ensure the consistency of storage-level logs (redo log and undo log) and server-level logs (binlog). This mechanism divides the transaction commit process into two phases: Prepare and Commit. The Commit phase is further subdivided into three stages: Flush, Sync, and Commit. 
In 2PL, a lock is required for each update, and these locks can be released after the commit phase in 2PC. Consequently, the order of the commit phase aligns with the order of updates, as depicted in Figure \ref{figue:design_group_lock_2pc_mysql}. However, in TXSQL with group locking, only the leader is required to acquire a lock, which may result in discrepancies between the update order and the order in which updates arrive during the commit phase. Therefore, it is essential to enforce consistency between the update order and the commit phase order. Furthermore, since the commit phase consists of three serially executed stages (i.e., Flush, Sync, and Commit), executing them strictly in accordance with the update order can easily lead to a critical path and performance bottleneck as depicted in Figure \ref{figue:design_group_lock_2pc_wo}.

To address this, we implement a group commit optimization in 2PC. 
As depicted in Figure \ref{figue:design_group_lock_2pc_w}, similar to group locking in 2PL, the first thread to enter the commit phase queue is designated as the leader, while subsequent threads function as followers. 
We ensure that transactions enter the Flush stage queue in the order of hotspot updates. When the leader of a commit phase group releases the lock, a group commit can occur. 
This approach not only guarantees the order of the commit phase but also leverages group commit to enhance overall efficiency.

\subsection{Rollback Order Guarantee}\label{sec:implementation_rollback_order}
TXSQL does not experience cascading abort in the absence of group locking. However, cascading aborts can occur in group locking, since the transaction can update the item without requiring a prior update to commit.
To ensure correct aborts, the rollback order for hotspot updates must align with the reverse update order specified in the dependency list.
The transaction rollback process, as shown in Algorithm \ref{alg:transaction_rollback}, involves setting transaction and item states, performing safety checks, and executing the rollback, as follows:

\begin{algorithm}[t]
    \caption{Transaction Rollback Process}
    \label{alg:transaction_rollback}
    \small
    \begin{algorithmic}[1]
        \Function{rollback}{T, item}
            \If{$T.\text{status} == \text{GRANTED}$}
                \State $item.\text{granting\_new\_trx} \gets \text{false}$
            \EndIf
            \If{$T.\text{hot\_update\_status}$ $==$ $\text{RUNNING} \land T.\text{is\_leader}$}
                \State $item.\text{switching\_new\_leader} \gets \text{false}$
            \EndIf
            
            \While{$(item.\text{dep\_list}.\text{back()}$ $\neq$ $trx$ $\lor$ $item.\text{granting\_new\_trx} \lor item.\text{switching\_new\_leader})$}
                \State ut\_delay(10) \Comment{\textcolor{gray!80}{Wait all subsequent transactions rollbacked}}
            \EndWhile
            
            \State rollback\_trx(T)
            \State $item.\text{dep\_list}.\text{erase}(T)$ \Comment{\textcolor{gray!80}{Remove from dependency list}}
        \EndFunction
    \end{algorithmic}
\end{algorithm}

Step 1 (Lines 2-5) - Set transaction and item states: If the transaction has been granted a hotspot update, the item will no longer grant new transactions. Additionally, if the hotspot update state of the transaction is RUNNING and the transaction is the leader, the item will not switch to a new leader.

Step 2 (Lines 6-7) - Check for safe rollback: Iteratively check whether the last transaction in the dependency list is not the current transaction and whether the item is granting new transactions or switching to a new leader. If either of these conditions is met, pause for a brief period before continuing the checks.

Step 3  (Lines 8-9) - Rollback the transaction: Execute the rollback of the transaction and remove it from the dependency list.

\noindent \textbf{Rollback optimization.} 
To achieve effective and efficient rollbacks,  it is essential to implement corresponding safeguards at both the Server and Storage layers. When the Server layer initiates a transaction rollback, it sends a signal to the Storage layer. Upon receiving this signal, the Storage layer refrains from granting new followers and does not release locks to new leaders. Our tests have shown that, compared to allowing continued updates on hotspots that require a rollback, this approach significantly reduces the number of transactions that may abort during the rollback period.
Once all transactions that need to be rolled back have been completed, another signal is sent to the Storage layer indicating that the rollback has finished, allowing it to resume granting and releasing operations. The Server layer is responsible for rolling back all uncommitted transactions but does not guarantee the order of the rollbacks; this order is ensured at the Storage layer. 
The dependency list allows us to determine whether a transaction is the most recent one to update the item. Only the transaction at the end of the list can be rolled back; other transactions must wait for the preceding transactions to complete their rollbacks.

\subsection{Deadlock Handling}\label{sec:implementation_deadlock_detection}
2PL can handle deadlocks through deadlock detection techniques or timeout mechanisms. When a deadlock is detected or a timeout occurs, the system can directly roll back one transaction to avoid deadlocks. However, with the introduction of our group locking mechanism in TXSQL, most updates occur without locks, making direct deadlock detection infeasible. 
An intuitive solution would be to integrate the dependency list into the existing deadlock detection mechanism of 2PL. However, this would inherently require more complex code logic. Additionally, as discussed in Section \ref{sec:design_queue_locking}, the effectiveness of deadlock detection is limited in scenarios involving hotspot access. Our preliminary experiments indicate that a timeout mechanism outperforms deadlock detection in terms of performance. However, the timeout rollback may lead to situations where transactions cannot be rolled back immediately, potentially resulting in excessively long chains of cascading rollbacks, which can significantly impact overall performance \cite{DBLP:conf/sigmod/GuoWYY21_BAMBOO}.

Consequently, we have redesigned our deadlock prevention mechanism. Deadlocks can be avoided through a timeout mechanism when accessing only non-hotspot rows, as this scenario does not lead to cascading aborts. Therefore, our focus is primarily on deadlocks that arise from simultaneous access to both hotspot and non-hotspot rows. When a blocking situation is detected, and both the current transaction and the blocking transaction have updated the same hotspot row, the likelihood of a deadlock occurring is considerably high. 
In such cases, we proactively initiate a rollback to prevent deadlocks. We have observed that this rollback incurs significantly less overhead than a timeout rollback.

\textcolor{black}{\noindent \textbf{Example of hotspot deadlock handling.} Assume the following scenario: the initial value of tuple t1 (id, val) = (1, 1), which is a hot row, and the initial value of tuple t2 (c1, c2) = (100, 100), which is a non-hot row. If both transactions first update the hot row and then update the non-hot row, the following situations may occur between them.}
\begin{table}[ht]
\centering
\setlength{\tabcolsep}{4pt}   
\renewcommand{\arraystretch}{1.2} 

\begin{tabular}{|p{0.21\textwidth}|p{0.21\textwidth}|} 
\hline
\textbf{Transaction 1} & \textbf{Transaction 2} \\
\hline
BEGIN; & \\
\hline
UPDATE t1 SET val = val + 1; & \\ 
  \textit{(hot row, updated to 2)} & 
BEGIN; \\
\hline
 & 
UPDATE t1 SET val = val + 1; \\ 
 & \textit{(hot row, success, val=3)} \\
\hline
 & 
UPDATE t2 SET c2 = c2 + 1; \\ 
  & \textit{(non-hot row, updated to 101)} \\
\hline
UPDATE t2 SET c2 = c2 + 1; & \\ 
  \textit{(non-hot row, blocked, waiting for T2's lock)} & \\
\hline
& COMMIT; \\ 
  & \textit{(blocked, depends on T1's commit, deadlock occurs, rollback)}  \\
\hline
\end{tabular}
\end{table}
\textcolor{black}{When transaction T2 is waiting for transaction T1's lock, if both T2 and T1 update the same hot row, a deadlock may occur regardless of which transaction goes first. If T2 updates the hot row first and T1 updates it afterward, the commit of T1 depends on T2, but T2 needs to wait for T1's lock, resulting in a deadlock. Conversely, if T1 updates the hot row first and T2 updates it afterward, although the order of commits does not cause a deadlock, the rollback of T1 depends on T2, which may also lead to a deadlock.}
\textcolor{black}{Therefore, when a transaction is blocked by a lock, if both the current transaction and the blocking transaction have updated the same hot row, the current locking step would be skipped and this transaction would be rolled back.}

\textcolor{black}{
\noindent \textbf{Example of hotspot cascade rollback.}
The tuple t1 (id, val) initially holds (1, 1) and becomes a hot row due to concurrent updates from three transactions: T1 modifies the value to 2; T3 subsequently updates it to 3; T2 further changes it to 4.}

\textcolor{black}{When T1 attempts to roll back, it encounters a dependency chain: since newer transactions (T3 and T2) have modified the hot row based on T1’s initial update, T1 must wait for these later transactions to roll back in reverse chronological order (T2 → T3). Only after all dependent transactions are rolled back can T1 safely revert its change.}

\begin{table}[ht]
\centering
\setlength{\tabcolsep}{4pt}   
\renewcommand{\arraystretch}{1.2} 

\begin{tabular}{|p{0.14\textwidth}|p{0.14\textwidth}|p{0.14\textwidth}|} 
\hline
\textbf{Transaction 1} & \textbf{Transaction 2} & \textbf{Transaction 3}\\
\hline
BEGIN; & & \\
\hline
 UPDATE t1 SET val = val + 1; & &\\
 \textit{(update to 2)} & &BEGIN; \\
\hline
  & & UPDATE t1 SET val = val + 1; \\
  & BEGIN; & \textit{(update to 3)}\\
\hline
 & UPDATE t1 SET val = val + 1;  &  \\
 & \textit{(update to 4)} & \\
\hline
ROLLBACK; & & \\
\textit{(wait until T2 and T3 rollback.)} & & \\
\hline
& COMMIT; & COMMIT; \\
& \textit{(rollback properly.)} & \textit{(wait until T2 rollback.)} \\
\hline
 & & COMMIT;\\
& & \textit{(rollback properly.)}  \\
\hline
ROLLBACK; & & \\
\textit{(rollback properly.)} & & \\
\hline
\end{tabular}
\end{table}


\textcolor{black}{
\subsection{Other Optimizations}\label{sec:implementation_group_lock_optimization} The group locking mechanism enhances performance for hotspot data updates, but it also introduces certain challenges in real-world applications. Next, we discuss these challenges and our optimizations related to group locking.
}

\subsubsection{Latency Optimization}\label{sec:implementation_fix_reqesut_latency_optimization}
Unlike academic research, which typically imposes no limits on transaction requests per second, the industry often employs a fixed Transactions Per Second (TPS) rate model. This model sends a predetermined number of transaction requests to the database each second, offering key advantages for customer workloads: (1) predictability in capacity planning and resource allocation, and (2) stability that ensures system availability under high load, minimizing the risk of crashes and performance degradation while maintaining consistent response times. 

In group locking, each group is assigned a specific batch size, and hotspot updates are managed within the granting and releasing processes of that group. However, under a fixed TPS rate model, the level of concurrency can fluctuate significantly, resulting in abrupt and temporary decreases in request volume. This can frequently lead to empty wait queues, causing a group to enter periods where no subsequent transactions are available for granting. As a result, these groups may remain indefinitely stalled while awaiting the next hotspot transaction, thereby increasing overall latency.

One simple approach is to reduce the batch size, thereby decreasing the probability of waiting. However, this strategy may lead to an increased number of groups, resulting in more frequent lock acquisitions, and does not effectively address the underlying issue of transaction waiting.
Another approach involves employing a background detection thread to periodically awaken the leader when possible. However, practical experience has shown that setting the period too short can lead to excessive checks of the $hot\_row\_hash$ table, resulting in an over-acquisition of mutexes and a significant decline in performance. Conversely, setting the period too long can still result in a substantial number of high-latency transactions.

To address these challenges, we ultimately adopted a dynamic batch size strategy. When no waiting transactions are available, the leader directly releases the lock without assigning it to a new leader. In this scenario, any upcoming hotspot update transactions can continue the role of leader and initiate a new group.





\textcolor{black}{
\subsubsection{Select for Update}
This optimization is primarily motivated by application requirements. Many workloads utilize the SELECT FOR UPDATE statement to lock records before executing updates. The coexistence of SELECT FOR UPDATE and UPDATE transactions introduces new challenges for hotspot locking, as both require the acquisition of locks. 
In general, an update statement should be followed by a SELECT FOR UPDATE statement. Typically in 2PL, the UPDATE and SELECT FOR UPDATE statements in a transaction should be completed in one time of locking without other update involved.
}

\textcolor{black}{
So our idea is also to ensure that the order in which transactions enter the queue for SELECT FOR UPDATE aligns with the order of updates and, ultimately, the sequence of transaction commits. In the implementation process, it is crucial to first verify whether a transaction is already queued when an update statement is received. If a SELECT FOR UPDATE statement for a transaction has previously queued, this transaction does not need to queue again, thereby reducing unnecessary waiting times.
Furthermore, it is essential to coordinate with the rollback sequence to maintain transactional integrity. 
Specifically, if a transaction rolls back after the SELECT FOR UPDATE but before the UPDATE, the states of siga and sigb must be accurately adjusted to reflect this rollback, ensuring that the system remains consistent and reliable.
}

\textcolor{black}{
\subsubsection{Replication Replay Optimization}
As discussed in Section \ref{sec:implementation_commit_order}, the native binlog operates independently for each transaction, requiring single-threaded replay. However, with group locking for hotspot data, binlogs can now support the 2PC group commit. This enhancement allows for multi-threaded replay on secondary servers, which we initially believed would yield improved performance. However, we observed that the parallel replay of these transactions led to significant lock contention, resulting in replay speeds that were even slower than those of single-threaded execution. This, in turn, caused excessive replication lag, adversely affecting the overall performance of the database.
}

\textcolor{black}{
To address this, we implemented a mechanism to ensure that if the current thread corresponds to a hotspot update transaction, it will not be replayed in parallel on the secondary server. 
This optimization enables us to reap the benefits of the group locking for hotspot data while avoiding the negative impacts of parallel replay on the secondary server.}

\section{Correctness}\label{sec:correctness}
This section discusses the correctness of transaction dependency, serializability, and failure recovery in the context of group locking. 

\subsection{Transaction Dependency}
The group locking mechanism enhances hotspot performance but compromises the original 2PL protocol, which requires a transaction to be committed before the next update begins. Group locking allows the next update to start immediately after completion, not after commit. To maintain the original execution and commit order of 2PL for hotspot data, three conditions must be met:
\begin{itemize}[left=0em]
\item Transaction commit order: When involving hotspot updates, a global queue by a dependency list is maintained to ensure that transactions are committed in the order of updates. Each transaction must verify that its predecessor has been committed before proceeding with its own commit.

\item Visibility of hotspot update: To ensure that the next hotspot update transaction can access the results of the preceding hotspot update, it is essential to refresh the update results immediately after each hotspot update transaction is complete. 

\item Transaction rollback order: When a transaction needs to be rolled back, all subsequent transactions that depend on it should be rolled back in reverse order according to the dependency list.
\end{itemize}


\subsection{Serializability}
According to the theory of serializability, a schedule of transactions is considered serializable if and only if its serialization graph is acyclic \cite{DBLP:journals/tse/BernsteinSW79,DBLP:conf/icde/AdyaLO00_adya}. In 2PL, every schedule is serializable, as forming a cyclic graph violates the two-phase locking rule \cite{DBLP:conf/sigmod/GuoWYY21_BAMBOO}. 
The core idea is that 2PL maintains the update order (i.e., conflict dependency in the conflict graph) consistent with the commit order. Therefore, to guarantee serializability in TXSQL, similar to Bamboo \cite{DBLP:conf/sigmod/GuoWYY21_BAMBOO} violating 2PL but guaranteeing serializability, it is essential to maintain consistency between the update order and the commit order.

The commit order can be strictly enforced by adhering to the update order based on the dependency list. However, in the initial design, parallel updates may occur during the leader-switching phases in group locking.
Consider a scenario in which transaction Trx1 is the last follower to be granted permission, while transaction Trx2 is the new leader that becomes awakened after the current leader releases the lock. Since Trx1 does not need to acquire any additional locks, it can directly update the hot row data. Simultaneously, Trx2 can also acquire the lock on the hotspot row and update the data. 
This situation lacks a mechanism to ensure the order of operations between these two transactions, potentially leading to an update loss anomaly, where both transactions read the same data and update it to different new values.

To prevent this, the old leader does not release the lock until all granted followers complete their updates, as discussed and implemented in Section \ref{sec:implementation_commit_order}.
By doing so, we ensure that the update order is serial and unique. Since the commit order can align with the update order, we conclude that TXSQL can achieve serializability with group locking.
Also, we describe the practical verification in the development process in Section \ref{sec:evaluation_correctness_check}.

\subsection{Failure Recovery}
In traditional 2PL failure recovery, unfinished transactions can roll back uncommitted updates individually, as the locking mechanism prevents concurrent transactions from simultaneously updating the same data row.
However, when dealing with hotspot updates through group locking, it is essential to roll back multiple unfinished transactions in the correct order.
To recap, the identifier $hot\_update\_order$, a globally incrementing identifier, has been added to the dependency list when hotspot updating. This identifier corresponds one-to-one with transactions, similar to the undo log header. We propose persisting the $hot\_update\_order$ within the undo log header, however, there are currently no reserved fields for this purpose.

After a transaction is committed, the $hot\_update\_order$ is removed from the dependency list and becomes ineffective, allowing us to repurpose the $TRX\_UNDO\_TRX\_NO$ field in the undo log header.
This field originally records the transaction's $trx\_no$, a sequence number generated upon commit to indicate the commit order of transactions. 
Since the effective periods of $hot\_update\_order$ and $trx\_no$ do not overlap, we designate the first bit of the field as 1 to indicate a $hot\_update\_order$; otherwise, it indicates a $trx\_no$.



Upon restart, we reconstruct the active transaction linked list from the undo log by reading the $TRX\_UNDO\_TRX\_NO$ field. If the transaction has not been committed, its globally incrementing value $hot\_update\_order$ can be restored, allowing us to reorder the active transaction linked list accordingly and proceed with sequential rollbacks in a single thread. 
If a failure occurs again during the rollback of hotspot transactions after a restart, those transactions that have already been rolled back will have completed their binlog entries. Consequently, upon the next restart, these transactions will not require rollback, allowing the remaining transactions to continue rolling back sequentially according to the $hot\_update\_order$, thereby maintaining correctness.
\pgfplotstableread{
1    5531.95  5607.78  5211.63  12466.21
2    4837.55  4983.77  4979     20835.79
3   3379.63  4110.56  4174.54  19286.86
4  2020.61  2933     3898.12  16670.38
}\fittps


\pgfplotstableread{
1	716.82	553.44	1740.28	1842.42
2	715.69	557.38	3981.16	4225.42
3	710.36	558.65	4562.61	7142.56
4	702.57	556.26	4249.54	10539.87
5	692.27	551.88	3726.03	14471.82
6	672.08	545.57	3328.17	14230.79
7	632.17	546.54	2720.33	13695.66
8	589.61	562.62	2009.89	13175.92
}\fitscalability

\pgfplotstableread{
1	11.45	92.42	5.2	4.91
2	22.69	132.49	3.68	3.49
3	45.79	204.11	7.98	5.47
4	92.42	511.33	15.83	7.04
5	186.54	1170.65	36.24	10.09
6	383.33	2082.91	82.96	19.65
7	816.63	3773.42	204.11	108.68
8	1771.29	5709.5	559.5	144.97
}\fitscalabilitylatency

\pgfplotstableread{
1	716.82	553.44	1740.28	1842.42
2	710.36	558.65	4562.61	7142.56
3	672.08	545.57	3328.17	14230.79
4	589.61	562.62	2009.89	13175.92
}\fitsynchronization

\pgfplotstableread{
1	5531.95	3015.4	11280.13	12466.21
2	4837.55	3108.18	9694.63	20835.79
3	3379.63	3490.58	5927.66	19286.86
4	2020.61	3996.05	2571.78	16670.38
}\fitasynchronization

\pgfplotstableread{
1	16896.64	19099.7
2	9257.28	10840.72
3	3628.49	4133.9
4	5665.45	14629.98
5	4469.91	9792.46
6	2389.21	3883.12
}\fitgroupcommit

\pgfplotstableread{
1	33.698	33.609	33.525	59.9952
2	33.147	32.944	33.089	111.84
3	34.35	31.818	31.898	107.0616
4	40.946	35.072	36.0192	106.2
}\fitcpu

\pgfplotstableread{
1	1.52	1.64	1.67	0.62
2	6.91	7.84	6.67	1.44
3	80.03	101.13	65.65	12.08
4	539.71	484.44	277.21	132.49
}\fitlatency
\pgfplotstableread{
1	1	1	1	1
2	6	4	4	1
3	71	60	55	13
4	512	336	259	63
}\fitlockwait

\pgfplotstableread{
index O1    O2    O3    O4
1	1	1	1	1
1	1.52	1.64	1.67	0.62
2	6	4	4	1
2	6.91	7.84	6.67	1.44
3	80.03	101.13	65.65	12.08
3	71	60	55	13
4	539.71	484.44	277.21	132.49
4	512	336	259	63
}\fitstackedlatency

\pgfplotstableread{
1	1.042393128	1	1.141958786	0.369569687
2	1.024347724	1	0.977893821	0.148136132
3	1.060924815	1	1.054168044	0.144032431
4	1.007550525	1	1.006553587	0.152641991
}\fitlocknumber

\pgfplotstableread{
index O1    O2    O3    O4
1	1.52	1.64	1.67	0.62
2	6.91	7.84	6.67	1.44
3	80.03	101.13	65.65	12.08
4	539.71	484.44	277.21	132.49
}\fitstackedlatencya
\pgfplotstableread{
index O1    O2    O3    O4
1	1	1	1	1
2	6	4	4	1
3	71	60	55	13
4	512	336	259	63
}\fitstackedlatencyb

\pgfplotstableread{
1	194964	181320	189030	214266
2	130782	166560	189120	207180
3	72450	56898	172578	202890
4	38064	50598	166080	178062
5	19464	42270	119148	121854
}\tpccwarehouse

\pgfplotstableread{
1	0.1409	0.1605	0.1063	0.095
2	0.2227	0.1704	0.1173	0.0947
3	0.4263	0.1815	0.1451	0.1267
4	0.8571	0.2979	0.1678	0.1484
5	1.7847	0.2881	0.2575	0.1844
}\tpccwarehousepaymentrt

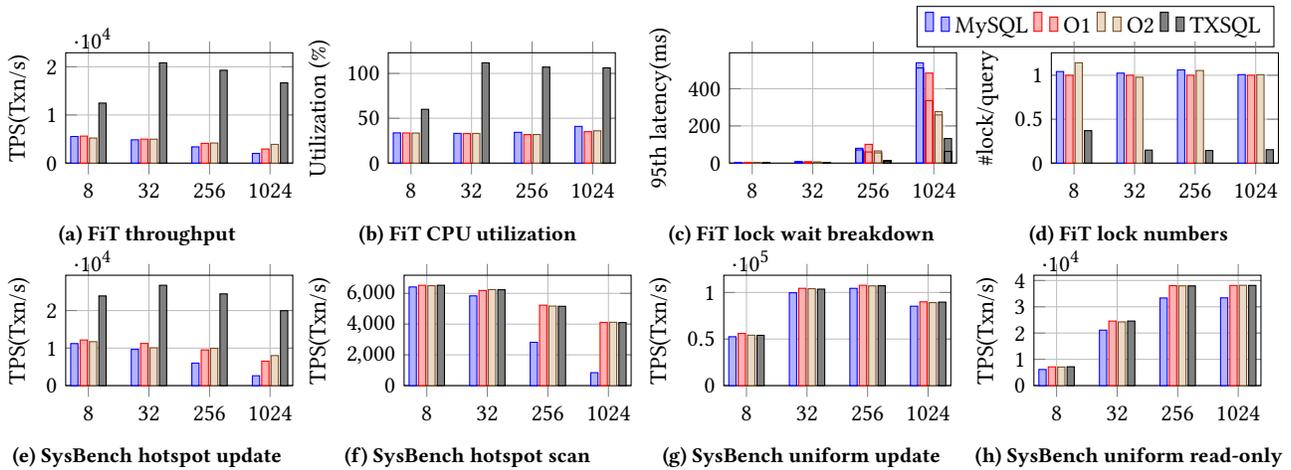
\begin{figure*}
    \centering
    \begin{subfigure}[b]{0.22\textwidth}
    \begin{tikzpicture}
        \begin{axis}[
            width=4.6cm,
            height=3.05cm,
            ylabel={TPS(Txn/s)},
            xtick={1,2,3,4},
            xticklabels={8,32,256,1024},
            ymin=0, 
            legend style={at={(1,1.05)}, anchor=south east},
            grid=major,
            bar width=0.12,
            enlarge x limits=0.125,
            ybar=2*\pgflinewidth,
            every ybar/.style={fill=blue!30},
            legend columns=4,
        ]

        \foreach \i in {1,2,3,4} {
            \addplot table[x index=0, y index=\i] {\fittps};
        }


        \end{axis}
    \end{tikzpicture}
    \caption{FiT throughput}\label{fig:effect_optimizations_fit_throughput}
    \end{subfigure}
    \begin{subfigure}[b]{0.25\textwidth}
    \begin{tikzpicture}
        \begin{axis}[
            width=4.6cm,
            height=3.05cm,
            ylabel={Utilization (\%)},
            xtick={1,2,3,4},
            xticklabels={8,32,256,1024},
            ymin=0, 
            legend style={at={(1,1.05)}, anchor=south east},
            grid=major,
            bar width=0.12,
            enlarge x limits=0.125,
            ybar=2*\pgflinewidth,
            every ybar/.style={fill=blue!30},
            legend columns=4,
        ]

        \foreach \i in {1,2,3,4} {
            \addplot table[x index=0, y index=\i] {\fitcpu};
        }

        \end{axis}
    \end{tikzpicture}
    \caption{FiT CPU utilization}\label{fig:effect_optimizations_cpu}
    \end{subfigure}
    \begin{subfigure}[b]{0.24\textwidth}
    \begin{tikzpicture}
        \begin{axis}[
            width=4.6cm,
            height=3.05cm,
            ylabel={95th latency(ms)},
            xtick={1,2,3,4},
            xticklabels={8,32,256,1024},
            ymin=0, 
            legend style={at={(2.4,1.05)}, anchor=south east},
            grid=major,
            bar width=0.12,
            enlarge x limits=0.125,
            ybar=2*\pgflinewidth,
            every ybar/.style={fill=blue!30},
            legend columns=4,
        ]

            \addplot table[x=index, y=O1] {\fitstackedlatency}; 
            \addplot table[x=index, y=O2] {\fitstackedlatency}; 
            \addplot table[x=index, y=O3] {\fitstackedlatency}; 
            \addplot table[x=index, y=O4] {\fitstackedlatency}; 

       \legend{MySQL,O1,O2,TXSQL}
        \end{axis}
    \end{tikzpicture}
    \caption{FiT lock wait breakdown}\label{fig:effect_optimizations_lock_wait}
    \end{subfigure}
    \begin{subfigure}[b]{0.24\textwidth}
    \begin{tikzpicture}
        \begin{axis}[
            width=4.6cm,
            height=3.05cm,
            ylabel={\#lock/query},
            xtick={1,2,3,4},
            xticklabels={8,32,256,1024},
            ymin=0, 
            legend style={at={(1,1.05)}, anchor=south east},
            grid=major,
            bar width=0.12,
            enlarge x limits=0.125,
            ybar=2*\pgflinewidth,
            every ybar/.style={fill=blue!30},
            legend columns=4,
        ]

        \foreach \i in {1,2,3,4} {
            \addplot table[x index=0, y index=\i] {\fitlocknumber};
        }


        \end{axis}
    \end{tikzpicture}
    \caption{FiT lock numbers}\label{fig:effect_optimizations_lock_number}
    \end{subfigure}
    \begin{subfigure}[b]{0.22\textwidth}
    \begin{tikzpicture}
        \begin{axis}[
            width=4.6cm,
            height=3.05cm,
            ylabel={TPS(Txn/s)},
            xtick={1,2,3,4},
            xticklabels={8,32,256,1024},
            ymin=0, 
            legend style={at={(1,1.05)}, anchor=south east},
            grid=major,
            bar width=0.12,
            enlarge x limits=0.125,
            ybar=2*\pgflinewidth,
            every ybar/.style={fill=blue!30},
            legend columns=4,
        ]

        \foreach \i in {1,2,3,4} {
            \addplot table[x index=0, y index=\i] {\sysbenchhotspot};
        }

        \end{axis}
    \end{tikzpicture}
    \caption{SysBench hotspot update}\label{fig:effect_optimizations_sys_hotspot_update}
    \end{subfigure}
    \begin{subfigure}[b]{0.25\textwidth}
    \begin{tikzpicture}
        \begin{axis}[
            width=4.6cm,
            height=3.05cm,
            ylabel={TPS(Txn/s)},
            xtick={1,2,3,4},
            xticklabels={8,32,256,1024},
            ymin=0, 
            legend style={at={(1,1.05)}, anchor=south east},
            grid=major,
            bar width=0.12,
            enlarge x limits=0.125,
            ybar=2*\pgflinewidth,
            every ybar/.style={fill=blue!30},
            legend columns=4,
        ]

        \foreach \i in {1,2,3,4} {
            \addplot table[x index=0, y index=\i] {\sysbenchscan};
        }

        \end{axis}
    \end{tikzpicture}
    \caption{SysBench hotspot scan}\label{fig:effect_optimizations_sys_hotspot_scan}
    \end{subfigure}
    \begin{subfigure}[b]{0.24\textwidth}
    \begin{tikzpicture}
        \begin{axis}[
            width=4.6cm,
            height=3.05cm,
            ylabel={TPS(Txn/s)},
            xtick={1,2,3,4},
            xticklabels={8,32,256,1024},
            ymin=0, 
            legend style={at={(1,1.05)}, anchor=south east},
            grid=major,
            bar width=0.12,
            enlarge x limits=0.125,
            ybar=2*\pgflinewidth,
            every ybar/.style={fill=blue!30},
            legend columns=4,
        ]

        \foreach \i in {1,2,3,4} {
            \addplot table[x index=0, y index=\i] {\sysbenchuniform};
        }

        \end{axis}
    \end{tikzpicture}
    \caption{SysBench uniform update}\label{fig:effect_optimizations_sys_uniform_update}
    \end{subfigure}
    \begin{subfigure}[b]{0.24\textwidth}
    \begin{tikzpicture}
        \begin{axis}[
            width=4.6cm,
            height=3.05cm,
            ylabel={TPS(Txn/s)},
            xtick={1,2,3,4},
            xticklabels={8,32,256,1024},
            ymin=0, 
            legend style={at={(1,1.05)}, anchor=south east},
            grid=major,
            bar width=0.12,
            enlarge x limits=0.125,
            ybar=2*\pgflinewidth,
            every ybar/.style={fill=blue!30},
            legend columns=4,
        ]

        \foreach \i in {1,2,3,4} {
            \addplot table[x index=0, y index=\i] {\sysbenchreadonly};
        }

        \end{axis}
    \end{tikzpicture}
    \caption{SysBench uniform read-only}\label{fig:effect_optimizations_sys_uniform_readonly}
    \end{subfigure}


    \caption{Effect of optimizations. The X-axis is the thread count, and the higher the more concurrent requests.}
    \label{fig:effect_optimizations}
\end{figure*}

\section{Evaluation}\label{Evaluation}
This section evaluates the performance of TXSQL empirically. 
Our goal is to validate three critical aspects of TXSQL: (1) its effectiveness in handling hotspot locks (Section~\ref{sec:evaluation_ablation_study}); (2) its performance superiority over state-of-the-art solutions across a variety of high-contention scenarios (Section~\ref{sec:evaluation_compare_sota_solutions}); (3) its performance under various workloads especially real-world ones in Tencent Cloud (Section~\ref{sec:evaluation_real_world_application}). 

\subsection{Setup}\label{sec:evaluation_setup}
We conducted our experiments on three servers, each equipped with an Intel(R) Xeon(R) Gold 6133 CPU @2.50GHz, featuring 80 cores, 753 GB of DRAM, and a 15 TB SSD. 
The operating system used was CentOS Linux release 7.2. The average network latency between servers was measured at 1.033 ms.

\subsubsection{Workloads} Four benchmarks were conducted as follows: 
(1) \textbf{SysBench~\cite{sysbench}} is a versatile open-source benchmarking tool that is frequently used in industrial database systems like MySQL and PostgreSQL. By default, the read/write (RW) ratio is set to 0.5, the transaction length (TL) is 14, and the default \textit{skew factor} (SF) is set to 0.7 for Zipf distribution. The main workloads include hotspot update (RW=0,TL=1; All updates in one hotspot row), hotspot mix read/write (RW=0.5,SF=0.9), hotspot scan (RW=0,TL=10), uniform update (RW=0,uniform), and uniform read-only (RW=1,uniform). 
(2) \noindent\textbf{TPC-C~\cite{DBLP:journals/pvldb/YangYHZYYCZSXYL22_oceanbase,DBLP:journals/pvldb/ChenPLYHTLCZD24_TDSQL}} is a popular OLTP benchmark that simulates e-commerce scenarios. 
The standard TPC-C is implemented and each warehouse contains about 100 MB of data, and we vary the number of warehouses to simulate different contentions.
(3) \noindent\textbf{FiT} simulates the transaction operations of the Financial Transaction system in Tencent. After data anonymization, the primary structure consists of two tables: a hot table that records account information with frequent updates to user balances, and a non-hot table that stores all transaction records. 
(4) \noindent\textbf{Hotspots} is a real-life workload combined with three applications that experience spikes of high-contented loads from time to time. 

\subsubsection{Baselines}
\textbf{MySQL:} The base database MySQL v8.0.30. 
\textbf{O1:} The general lock optimization discussed in Section \ref{sec:design_general_lock_optimization}. 
\textbf{O2:} O1 and the queue locking optimization discussed in Section \ref{sec:design_queue_locking}. 
\textbf{TXSQL:} O1 and the group locking optimization discussed in Section \ref{sec:design_group_locking}. By default, the group locking batch size is set to 10. 
\textbf{Aria \cite{DBLP:journals/pvldb/LuYCM20_aria}:} A state-of-the-art (SOTA) deterministic protocol representative which combines OCC and deterministic protocol without pre-knowing read-write sets. We tuned the best batch size for Aria.
\textbf{Bamboo \cite{DBLP:conf/sigmod/GuoWYY21_BAMBOO}:} SOTA solution that violates 2PL to boost perofrmance.
For an apples-to-apples comparison, we implemented all approaches, including Aria and Bamboo, within TXSQL.

\subsection{Ablation Study}\label{sec:evaluation_ablation_study}
This part evaluates the effectiveness of the optimizations in TXSQL against MySQL, general lock optimization (\textbf{O1}), and queue locking optimization (\textbf{O2}) using FiT and SysBench workloads. 

\subsubsection{In-depth Performance Analysis}
In the FiT workload, as the number of threads increases, both O1 and O2 show improvements in lock waiting times compared to MySQL, as shown in Figure \ref{fig:effect_optimizations_lock_wait}, as they reduce the lock-holding time, however, the 95th percentile latency is slightly higher in O1 due to the overhead required for actively detecting deadlocks (The entire bar indicates the transaction latency, while the line within the bar represents the lock waiting time). The benefits of O2 are still limited due to its long lock wait time for the hotspot. TXSQL, through the implementation of group locking, significantly reduces the number of locks, as shown in Figure \ref{fig:effect_optimizations_lock_number}. As a result, TXSQL achieves better CPU utilization, as shown in Figure \ref{fig:effect_optimizations_cpu}, resulting in a clear throughput advantage over MySQL, O1, and O2, with improvements of up to 8.25x, as shown in Figure \ref{fig:effect_optimizations_fit_throughput}. 

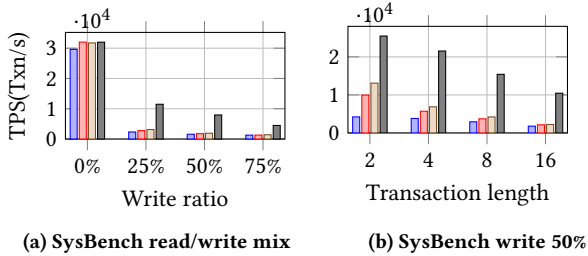
\begin{figure}
    \centering
    \begin{subfigure}[b]{0.235\textwidth}
    \begin{tikzpicture}
        \begin{axis}[
            width=4.5cm,
            height=3cm,
            xlabel={Write ratio},
            ylabel={TPS(Txn/s)},
            xtick={1,2,3,4},
            xticklabels={0\%,25\%,50\%,75\%},
            ymin=0, 
            legend style={at={(1,1.05)}, anchor=south east},
            grid=major,
            bar width=0.12,
            enlarge x limits=0.125,
            ybar=2*\pgflinewidth,
            every ybar/.style={fill=blue!30},
            legend columns=4,
        ]

        \foreach \i in {1,2,3,4} {
            \addplot table[x index=0, y index=\i] {\sysbenchreadwriteratio};
        }

        \end{axis}
    \end{tikzpicture}
    \caption{SysBench read/write mix}
    \end{subfigure}
    \begin{subfigure}[b]{0.235\textwidth}
    \begin{tikzpicture}
        \begin{axis}[
            width=4.5cm,
            height=3cm,
            xlabel={Transaction length},
            xtick={1,2,3,4},
            xticklabels={2,4,8,16},
            ymin=0, 
            legend style={at={(1,1.05)}, anchor=south east},
            grid=major,
            bar width=0.12,
            enlarge x limits=0.125,
            ybar=2*\pgflinewidth,
            every ybar/.style={fill=blue!30},
            legend columns=4,
        ]

        \foreach \i in {1,2,3,4} {
            \addplot table[x index=0, y index=\i] {\sysbenchlength};
        }

        \end{axis}
    \end{tikzpicture}
    \caption{SysBench write 50\%}\label{fig:effect_optimizations2_transcation_length}
    \end{subfigure}
    \caption{Effect of optimizations (Thread=1024). }
    \label{fig:effect_optimizations2}
\end{figure}

\subsubsection{Hotspot and Non-hotspot Scenarios}
In SysBench workloads, TXSQL achieved up to a 7.5x improvement in hotspot update workload compared to those observed in the FiT workload, as shown in Figure \ref{fig:effect_optimizations_sys_hotspot_update}. In uniform scenarios (both updates and read-only workloads), as shown in Figures \ref{fig:effect_optimizations_sys_uniform_update} and \ref{fig:effect_optimizations_sys_uniform_readonly}, O2 and TXSQL show no performance improvement over O1, as neither triggers a hotspot lock mechanism. Luckily, the overhead of hotspot detection for O2 and TXSQL remained below 2\% in non-hotspot scenarios. In the hotspot scan workload, as shown in Figure \ref{fig:effect_optimizations_sys_hotspot_scan}, O2 and TXSQL also showed no improvement due to the dispersion of updates across multiple hotspots. O2 and TXSQL implemented optimizations of O1, resulting in performance improvements over MySQL in general non-hotspot scenarios.

To conduct a more thorough analysis, we varied the write ratio and transaction length in SysBench hotspot update (TL=20) scenarios, as shown in Figure \ref{fig:effect_optimizations2}. TXSQL achieved the best performance across all scenarios, despite a decline in performance across all systems with increasing write ratios or transaction lengths. Notably, O2 is somewhat effective when the transaction length is small compared to MySQL; however, its effectiveness diminishes with larger transaction lengths. This is because the increased latency exacerbates the lock contention in hotspot scenarios. 

\subsection{Comparision to State-of-the-art Solutions}\label{sec:evaluation_compare_sota_solutions}
This part evaluates the efficiency of TXSQL against \textbf{MySQL}, a SOTA deterministic protocol (\textbf{Aria}), and a SOTA 2PL protocol (\textbf{Bamboo}) by various scenarios.

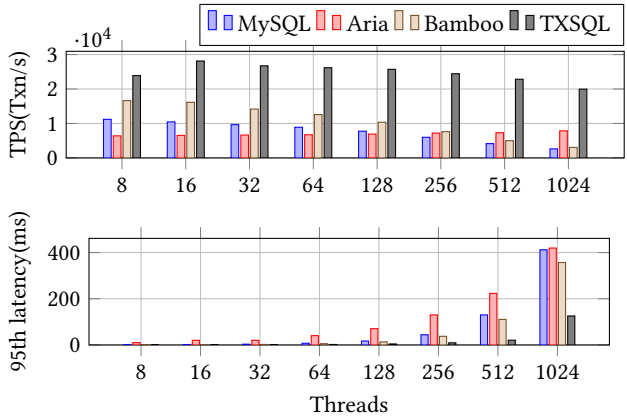
\begin{figure}
    \centering
    \begin{subfigure}[b]{0.5\textwidth}
    \begin{tikzpicture}
        \begin{axis}[
            width=9cm,
            height=3cm,
            ylabel={TPS(Txn/s)},
            xtick={1,2,3,4,5,6,7,8},
            xticklabels={8,16,32,64,128,256,512,1024},
            ymin=0, 
            legend style={at={(1,1.05)}, anchor=south east},
            grid=major,
            bar width=0.12,
            enlarge x limits=0.125,
            ybar=2*\pgflinewidth,
            every ybar/.style={fill=blue!30},
            legend columns=4,
        ]

        \foreach \i in {1,2,3,4} {
            \addplot table[x index=0, y index=\i] {\sysbenchscalability};
        }

        \legend{MySQL,Aria,Bamboo,TXSQL}
        \end{axis}
    \end{tikzpicture}
    \end{subfigure}
    \begin{subfigure}[b]{0.5\textwidth}
    \begin{tikzpicture}
        \begin{axis}[
            width=8.5cm,
            height=3cm,
            xlabel={Threads},
            ylabel={95th latency(ms)},
            xtick={1,2,3,4,5,6,7,8},
            xticklabels={8,16,32,64,128,256,512,1024},
            ymin=0, 
            legend style={at={(1,1.05)}, anchor=south east},
            grid=major,
            bar width=0.12,
            enlarge x limits=0.125,
            ybar=2*\pgflinewidth,
            every ybar/.style={fill=blue!30},
            legend columns=4,
        ]

        \foreach \i in {1,2,3,4} {
            \addplot table[x index=0, y index=\i] {\sysbenchscalabilitylatency};
        }

        \end{axis}
    \end{tikzpicture}
    \end{subfigure}
    \caption{Effect of scalability by Sysbench hotspot update. }
    \label{fig:effect_scalability}
\end{figure}

\subsubsection{Scalability}
We evaluated scalability using the SysBench hotspot update, as illustrated in Figure \ref{fig:effect_scalability}. With an increase in the number of threads, TXSQL achieved a performance improvement of up to 7x. This is attributed to the group locking mechanism, which significantly reduces lock contention for transactions at hotspots, thereby decreasing overall transaction latency, as evidenced by the 95th percentile latency statistics.
Bamboo demonstrated superior performance under low concurrency compared to MySQL and Aria; however, its performance improvement under high concurrency was less effective. This is primarily because, although locks can be released before the transaction commits and the lock-holding time remains relatively short compared to Aria and MySQL,  each update still requires acquiring locks, leading to substantial lock contention at hotspots. 
Aria, utilizing a deterministic algorithm, maintained stable TPS as the number of threads increased, thanks to its consistent scheduling of batch-size transactions.

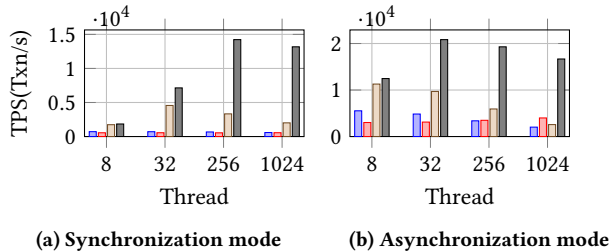
\begin{figure}
    \centering
    \begin{subfigure}[b]{0.235\textwidth}
    \begin{tikzpicture}
        \begin{axis}[
            width=4.5cm,
            height=3cm,
            xlabel={Thread},
            ylabel={TPS(Txn/s)},
            xtick={1,2,3,4},
            xticklabels={8,32,256,1024},
            ymin=0, 
            legend style={at={(1,1.05)}, anchor=south east},
            grid=major,
            bar width=0.12,
            enlarge x limits=0.125,
            ybar=2*\pgflinewidth,
            every ybar/.style={fill=blue!30},
            legend columns=4,
        ]

        \foreach \i in {1,2,3,4} {
            \addplot table[x index=0, y index=\i] {\fitsynchronization};
        }

        \end{axis}
    \end{tikzpicture}
    \caption{Synchronization mode}
    \end{subfigure}
    \begin{subfigure}[b]{0.235\textwidth}
    \begin{tikzpicture}
        \begin{axis}[
            width=4.5cm,
            height=3cm,
            xlabel={Thread},
            xtick={1,2,3,4},
            xticklabels={8,32,256,1024},
            ymin=0, 
            legend style={at={(1,1.05)}, anchor=south east},
            grid=major,
            bar width=0.12,
            enlarge x limits=0.125,
            ybar=2*\pgflinewidth,
            every ybar/.style={fill=blue!30},
            legend columns=4,
        ]

        \foreach \i in {1,2,3,4} {
            \addplot table[x index=0, y index=\i] {\fitasynchronization};
        }

        \end{axis}
    \end{tikzpicture}
    \caption{Asynchronization mode}
    \end{subfigure}
    \caption{Effect of synchronization by FiT.}
    \label{fig:effect_synchronization}
\end{figure}

\subsubsection{Synchronization}
In scenarios involving both synchronous and asynchronous replication, as illustrated in Figure \ref{fig:effect_synchronization}, TXSQL achieved performance improvements of 22.3x and 8.2x, respectively. Although the overall performance declined with synchronous replication, TXSQL demonstrated a relatively greater enhancement. This is attributed to the longer lock-holding times of transactions, which the group locking mechanism effectively mitigates. This phenomenon is akin to the situation where an increase in transaction length leads to performance degradation, as shown in Figure \ref{fig:effect_optimizations2_transcation_length}.
Bamboo performs well under low concurrency due to its ability to release locks earlier, thereby minimizing conflicts. However, as contention increases, the intensification of lock competition results in suboptimal performance. In the context of synchronous replication, Bamboo outperforms both Aria and MySQL, as the higher transaction latency in synchronous scenarios means that early lock release has a direct impact on reducing latency and enhancing performance.


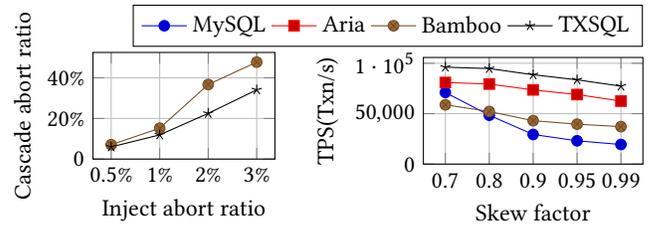
\begin{figure}
    \centering
    \begin{subfigure}[b]{0.09\textwidth}
    \begin{tikzpicture}
        \begin{axis}[
            width=4cm,
            height=3cm,
            xlabel={Inject abort ratio},
            ylabel={Cascade abort ratio},
            xtick={2,3,4,5},
            xticklabels={0.5\%,1\%,2\%,3\%},
            ytick={0,0.2,0.4},
            yticklabels={0,20\%,40\%},
            ymin=0, 
            legend style={at={(1,1.05)}, anchor=south east},
            grid=major,
            enlarge x limits=0.125,
            legend columns=4,
        ]

        \addplot table[x index=0, y index=1] {\sysbenchabortrate}; 
        \addplot table[x index=0, y index=2] {\sysbenchabortrate}; 
        \addplot table[x index=0, y index=3] {\sysbenchabortrate};
        \addplot table[x index=0, y index=4] {\sysbenchabortrate};

        \end{axis}
    \end{tikzpicture}
    \end{subfigure}
    \begin{subfigure}[b]{0.375\textwidth}
    \begin{tikzpicture}
        \begin{axis}[
            width=4.5cm,
            height=3cm,
            xlabel={Skew factor},
            ylabel={TPS(Txn/s)},
            xtick={1,2,3,4,5},
            xticklabels={0.7,0.8,0.9,0.95,0.99},
            ymin=0, 
            legend style={at={(1,1.05)}, anchor=south east},
            grid=major,
            enlarge x limits=0.125,
            legend columns=4,
            legend style={at={(1,1.1)}, anchor=south east}, 
            y tick label style={scaled ticks=base 10:0},
        ]

        \addplot table[x index=0, y index=1] {\sysbenchskewness}; \addlegendentry{MySQL}
        \addplot table[x index=0, y index=2] {\sysbenchskewness};\addlegendentry{Aria}
        \addplot table[x index=0, y index=3] {\sysbenchskewness}; \addlegendentry{Bamboo}
        \addplot table[x index=0, y index=4] {\sysbenchskewness}; \addlegendentry{TXSQL}

        \end{axis}
    \end{tikzpicture}
    \end{subfigure}
    \caption{Effect of abort and skewness by SysBench.}
    \label{fig:effect_abort_skewness}
\end{figure}

\subsubsection{Effect of Abort} This part tests the impact of cascading rollbacks on TXSQL and Bamboo by actively injecting rollback transactions during the SysBench hotspot update (RW=0.5, TL=16). As shown in Figure \ref{fig:effect_abort_skewness} (left), TXSQL outperformed Bamboo in terms of handling cascading rollbacks; however, both systems experienced significant effects, with rollback ratios increasing by more than tenfold, leading to substantial performance degradation. Cascading rollbacks were more likely to occur when transaction lengths were high and multiple hotspots were present. Luckily, for Tencent Cloud applications, we did not observe significant rollbacks when enabling group locking, as most transactions have relatively short lengths \cite{DBLP:conf/sosp/TuZKLM13,DBLP:journals/pvldb/KallmanKNPRZJMSZHA08-hstore,DBLP:conf/osdi/Lu00L23_NCC} and typically involve only one hotspot.
Additionally, we can monitor the rollback rate, and if it becomes excessively high, we can disable group locking and revert to the original 2PL protocol on the fly.

\subsubsection{Skewness}
This part tests the impact of access skewness using SysBench update workload (TL=1), as shown in Figure \ref{fig:effect_abort_skewness} (right). As skewness increases, contention rises, leading to a significant enhancement in TXSQL's performance, with improvements ranging from 1.6x to 3.9x. Notably, despite the high skewness, the contention level does not exceed that of a hotspot workload. In scenarios with a skewness level of 0.99, multiple hotspots may arise; however, the queuing situation for these hotspots is less severe than the contention observed during our previous tests with the SysBench hotspot update involving a single hotspot.
The limited improvement observed in Bamboo can be attributed to the relatively short duration of the workload transactions, resulting in a shorter overall lock-holding time and minimal optimization potential. Conversely, Aria experiences a gradual increase in the impact of rollbacks on performance as skewness rises, with a rollback rate exceeding 20\% at a skewness level of 0.99.

\begin{figure} 
\centering
\includegraphics[width=0.48\textwidth]{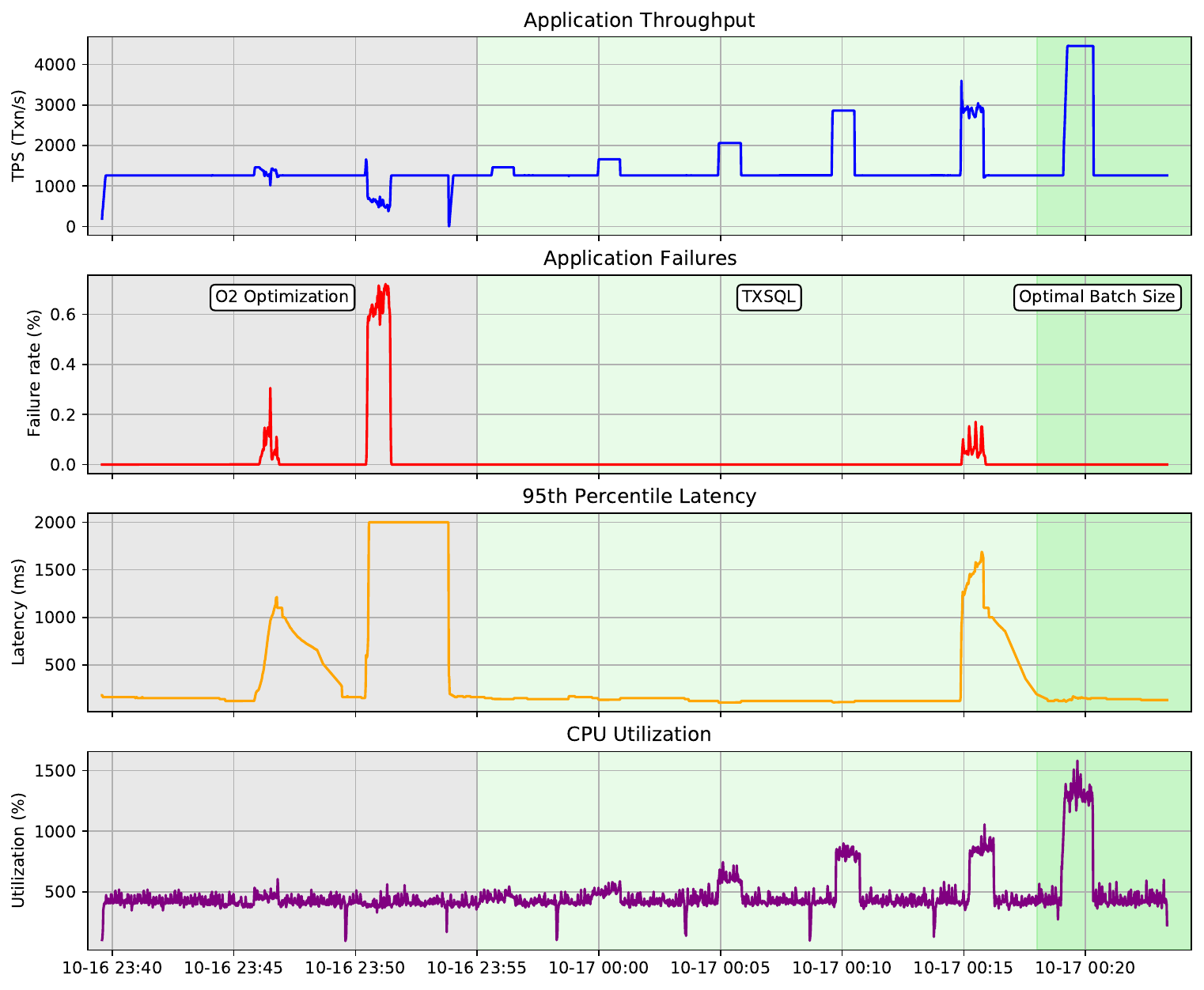}
\caption{Online application workloads with hotspots.}
\label{fig:evaluation_online_application}
\end{figure}

\subsection{More Workloads and Scenarios}\label{sec:evaluation_real_world_application}
\subsubsection{Real-world Applications}
Since 2023, our internal FIT payment and financial services have upgraded over 20,000 instances from MySQL to the TXSQL kernel. This transition has resulted in an overall performance improvement of 30\% and has collaboratively addressed long-standing performance jitter issues, where sudden spikes of load requests may significantly decrease the throughput. Additionally, the enhanced capability for handling hotspot update transactions has boosted performance by nearly 10 times. 

We next present a real-world hotspot workload, a composite of three applications in Tencent Cloud, adhering to a fixed TPS rate, as shown in Figure \ref{fig:evaluation_online_application}. 
It is important to note that the machines employed in this context are distinct from those utilized in other experiments. In order to maintain consistency with the online instance deployment, the specific configuration adopted here consists of an Intel(R) Xeon(R) CPU E5-2620 24 cores, 128 GB DRAM, 1.8 TB SSD, with an average network latency of 2.516 ms between the source and 2 semi-sync replicas.
The group locking optimization is an on-the-fly configurable parameter, activated after 23:55. We observe that, for the majority of the time, the TPS remains at a stable level. However, during the period when group locking is disabled, the system's processing capacity is adversely affected by sudden surges in request volume, leading to increased failure rates and latency. For instance, at 23:52, a hotspot emerges, causing the TPS to drop below its original level, despite CPU utilization remaining low. 
Once group locking is enabled in TXSQL, the system effectively manages sudden requests with minimal transaction failures and latency increases. However, as TPS continues to rise under sustained request conditions, TXSQL can process higher throughput but may experience some degree of failure rates and increased latency. Nevertheless, by further increasing the batch size within the group at 00:18, TXSQL is still able to handle hotspots efficiently.


\begin{figure}
    \centering
    \begin{subfigure}[b]{0.12\textwidth}
    \begin{tikzpicture}
        \begin{axis}[
            width=4.5cm,
            height=3cm,
            xlabel={\#Warehouse},
            ylabel={TPS(Txn/s)},
            xtick={1,2,3,4,5},
            xticklabels={16,8,4,2,1},
            ymin=0, 
            legend style={at={(1,1.05)}, anchor=south east},
            grid=major,
            bar width=0.12,
            enlarge x limits=0.125,
            ybar=2*\pgflinewidth,
            every ybar/.style={fill=blue!30},
            legend columns=4,
        ]

        \foreach \i in {1,2,3,4} {
            \addplot table[x index=0, y index=\i] {\tpccwarehouse};
        }

        \end{axis}
    \end{tikzpicture}
    \end{subfigure}
    \begin{subfigure}[b]{0.345\textwidth}
    \begin{tikzpicture}
        \begin{axis}[
            width=4.5cm,
            height=3cm,
            xlabel={\#Warehouse},
            ylabel={Latency(s)},
            xtick={1,2,3,4,5},
            xticklabels={16,8,4,2,1},
            ymin=0, 
            legend style={at={(1,1.05)}, anchor=south east},
            grid=major,
            bar width=0.12,
            enlarge x limits=0.125,
            ybar=2*\pgflinewidth,
            every ybar/.style={fill=blue!30},
            legend columns=4,
        ]

        \foreach \i in {1,2,3,4} {
            \addplot table[x index=0, y index=\i] {\tpccwarehousepaymentrt};
        }

        \legend{MySQL,Aria,Bamboo,TXSQL}
        \end{axis}
    \end{tikzpicture}
    \end{subfigure}
    \caption{Effect of warehouses by TPC-C.}
    \label{fig:effect_tpcc_warehouse}
\end{figure}
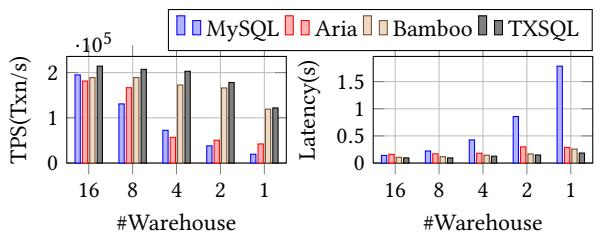

\subsubsection{TPC-C}
Under TPC-C workloads, a smaller number of warehouses leads to a greater number of conflicts. When the number of warehouses is set to one, TXSQL demonstrates a relatively better performance improvement compared to other opponents, as shown in Figure \ref{fig:effect_tpcc_warehouse} (left). However, Bamboo also performs well under this workload, primarily because TPC-C workloads consist of long transactions, which benefit significantly from early lock releases. Nevertheless, Bamboo incurs the overhead of locking and unlocking operations for each SQL statement, which is more costly than group locking. In the case of Aria, although latency is low with fewer warehouses (as indicated by the average latency of Payment transactions in Figure \ref{fig:effect_tpcc_warehouse} (right)), the high rate of transaction rollbacks negatively impacts its overall performance.

\begin{figure}
    \centering
    \begin{subfigure}[b]{0.235\textwidth}
    \begin{tikzpicture}
        \begin{axis}[
            width=4.5cm,
            height=3cm,
            xlabel={Batch size},
            ylabel={TPS(TXN/s)},
            xtick={1,2,3,4,5},
            xticklabels={1,4,16,64,256},
            ytick={0,10000,20000},
            yticklabels={0,10k,20k},
            ymin=0, 
            legend style={at={(1,1.05)}, anchor=south east,font=\footnotesize},
            grid=major,
            enlarge x limits=0.125,
            legend columns=2,
            y tick label style={scaled ticks=base 10:0},
        ]

        \addplot table[x index=0, y index=1] {\sysbenchbatchsizeone}; \addlegendentry{FIT-512}
        \addplot table[x index=0, y index=1] {\sysbenchbatchsizetwo}; \addlegendentry{FIT-32}
        \addplot table[x index=0, y index=3] {\sysbenchbatchsizeone}; \addlegendentry{ HRW-512}
        \addplot table[x index=0, y index=3] {\sysbenchbatchsizetwo}; \addlegendentry{ HRW-32}
        \addplot table[x index=0, y index=2] {\sysbenchbatchsizeone}; \addlegendentry{HU-512}
        \addplot table[x index=0, y index=2] {\sysbenchbatchsizetwo}; \addlegendentry{HU-32}

        \end{axis}
    \end{tikzpicture}
    \end{subfigure}
    \begin{subfigure}[b]{0.235\textwidth}
    \begin{tikzpicture}
        \begin{axis}[
            width=4.5cm,
            height=3cm,
            xtick={1,2,3,4,5,6},
            xticklabels={FIT-AS,HRW-AS,HU-AS,FIT-S,HRW-S,HU-S},
            xticklabel style={rotate=75}, 
            ytick={0,10000,20000},
            yticklabels={0,10k,20k},
            ymin=0, 
            legend style={at={(1,1.05)}, anchor=south east},
            grid=major,
            bar width=0.12,
            enlarge x limits=0.125,
            ybar=2*\pgflinewidth,
            every ybar/.style={fill=blue!30},
            legend columns=4,
        ]

        \foreach \i in {1,2} {
            \addplot table[x index=0, y index=\i] {\fitgroupcommit};
        }

        \legend{W/O GC, W GC}
        \end{axis}
    \end{tikzpicture}
    \end{subfigure}
    \caption{Effect of batch size and group commit (GC). }
    \label{fig:effect_batch_size_group_commit}
\end{figure}
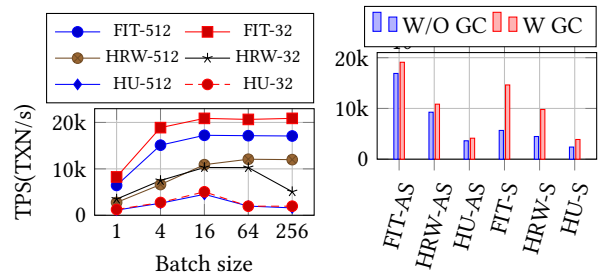

\subsubsection{Batch Size}
We investigated the impact of fixed batch size for group locking on performance with two thread settings (512 and 32), as shown in Figure \ref{fig:effect_batch_size_group_commit}.  Initially, performance improved with larger batch sizes due to fewer required locks. In shorter transactions, like FIT, performance remained stable even at a batch size of 256. However, in longer update transactions, such as the SysBench hotspot update (HU) workload (RW=0,TL=16), performance declined beyond a certain batch size due to increased conflicts and rollbacks.
Interestingly, in the SysBench hotspot read/write (HRW) workload (RW=0.5,TL=16), performance with large batch sizes was poorer under low concurrency (thread=32). Here, transaction arrival rates were lower than the rate at which group locking activated transactions, necessitating background threads to periodically check for waiting transactions. This indicates that maintaining a fixed batch size under a fixed TPS rate can increase latency and degrade performance. As discussed in Section \ref{sec:implementation_fix_reqesut_latency_optimization}, a dynamic batch size strategy can mitigate this issue by allowing the leader to release locks directly when no waiting transactions are present. 

\subsubsection{Group Commit}
We continued the workloads from the previous subsection to test the performance comparison of group commit under both synchronous (S) and asynchronous (AS) modes with a setting of 512 threads. As shown in Figure \ref{fig:effect_batch_size_group_commit}, we observe a greater benefit from group commit in synchronous mode. This is attributed to the increased occurrence of grouped transactions, which enhances the likelihood of reducing network communication between the synchronization and commit phases.


\subsubsection{Correctness Check} \label{sec:evaluation_correctness_check} We employed various practical testing methods to ensure the correctness of results produced by the group locking mechanism. For specific workloads, we verify consistency by checking if the results align with the expected outcomes based on logical operations. For instance, in TPC-C, we confirm that total sales for a Warehouse match those of its District. In FiT, we log each transaction and compare the total transaction value with recorded values. We also use database and business logs in a reconciliation system for verification.
Additionally, we utilize transaction correctness validation tools, including our DBChaos testing platform, which tests the execution results under numerous fault scenarios, including infrastructure failures (such as disk and network) and custom faults. Before going live, we conduct a \textit{canary} release using two parallel systems: one with group locking deployed and the one without, performing data verification every minute. We also conduct isolation checks using tools such as Jepsen \cite{DBLP:journals/pvldb/AlvaroK20_elle} and IsoVista \cite{DBLP:journals/pvldb/GuLXWCB24_isovista} to ensure the isolation correctness of our implementation.



\subsubsection{Failure Recovery} 
This part evaluated the restart duration by terminating processes and reconnecting to the client.
In the scenarios involving FiT-512, HRU-512, and HU-512, TXSQL achieved a TPS improvement of 5x to 13x compared to MySQL. The crash recovery durations were approximately 10 seconds for TXSQL, 7 seconds for MySQL, and 5 seconds for TXSQL without active transactions. The longer recovery time for TXSQL is due to both the length of active transactions and the volume of redo logs that need to be applied, which increases with higher TPS.

\subsection{Insight and Discussion}
\noindent \textbf{Potential for performance improvement.} 
This paper focuses on improving lock contention in concurrency control protocols for hotspot scenarios. In such scenarios, CPU and memory resources are not the bottlenecks; rather, the contention arises from single data access points. There are other avenues to address hotspot issues that complement group locking optimization. For instance, caching hotspot row locations in the index can mitigate overhead from repeated index lookups \cite{DBLP:journals/pvldb/HaoC24}. 
Additionally, we utilize a 2PC mechanism for both binlog and redolog to ensure the consistency of logical and physical logs, thus maintaining data correctness in extreme situations. However, this approach can significantly affect performance. Since the contents of these logs are consistent, we can optimize the mechanism by merging them or converting one into the other, allowing for a single log retention \cite{DBLP:journals/pvldb/ChenPLYHTLCZD24_TDSQL}.

\noindent \textbf{Trade-offs for simplicity and generality.} 
The current solution is limited to updating a single hotspot row within a transaction and does not support transactions involving multiple hotspot rows. Concurrent transactions involving multiple hotspot rows can easily lead to deadlocks, causing cascading rollbacks and negatively impacting performance. In practice, transactions typically update only one hotspot at a time. If two rows within the same transaction are frequently updated simultaneously, it is advisable to merge these data elements into the same table or row according to database normalization principles (e.g., 3NF), to reduce data redundancy and enhance consistency. Our solution was compared only with one system MySQL for fairness. However, the group locking should be easily and effectively applicable to other 2PL systems.

\noindent \textbf{Statements and isolation levels.} 
UPDATE, INSERT, and DELETE statements, which are classified as Current Reads collectively referred to as update operations, require locks. Common SELECT statements under Repeatable Read and Read Committed isolation levels are considered Snapshot Reads and do not require locks. However, under the Serializable (SER) isolation level, SELECT is treated as a Current Read and requires locks. While the lock optimization in this paper is independent of isolation levels, this paper focuses on the strongest SER levels and supports SELECT FOR UPDATE. Since INSERT and DELETE operations do not create hotspots, \textit{Phantom Reads} can be avoided within the group locking mechanism through next-key locking in indexes \cite{DBLP:conf/vldb/Mohan90_next_key}.


\section{Related Work}\label{sec:related_work}
\noindent \textbf{OLTP optimizations.}
The recent researched concurrency control protocols, since 2012, can be classified into six categories: two-phase locking (2PL) \cite{DBLP:journals/pvldb/RenTA12_vll,DBLP:conf/sigmod/GuoWYY21_BAMBOO}, timestamp ordering (TO), multi-version concurrency control (MVCC) \cite{DBLP:conf/eurosys/YabandehF12,DBLP:journals/pvldb/FaleiroA15,DBLP:conf/sigmod/0001MK15,durner2019no}, optimistic concurrency control (OCC) \cite{DBLP:conf/sosp/TuZKLM13,DBLP:journals/pvldb/YuanWLDXBZ16,DBLP:conf/sigmod/YuPSD16_tictoc,DBLP:journals/pvldb/YuXPSRD18}, deterministic concurrency control \cite{DBLP:conf/sigmod/ThomsonDWRSA12_calvin,DBLP:conf/osdi/MuCZLL14,DBLP:journals/pvldb/FaleiroAH17,DBLP:journals/pvldb/LuYCM20_aria,DBLP:journals/jcst/DongTWWCZ20_DOCC,DBLP:conf/sigmod/LinTLCW21_Hermes,DBLP:conf/sosp/QinBG21_Caracal,DBLP:conf/sigmod/BoeschenB22_GaccO}, and adaptive concurrency control \cite{DBLP:conf/sosp/XieSLAK015_Callas,DBLP:conf/sigmod/ShangLYZC16_HSync,DBLP:conf/cidr/TangJE17,DBLP:conf/sigmod/SuCDAX17_Tebaldi,DBLP:conf/usenix/TangE18_CormCC,DBLP:conf/icde/Su0Z21_C3,DBLP:conf/osdi/WangDWCW0021}. 
At present, deterministic and adaptive protocols are progressively gaining prominence in academia. However, deterministic protocols require prior knowledge of the read-write sets, while adaptive protocols pose significant challenges to correctness. As a result, these protocols are limited in adoption in the industry. 
This paper mainly focuses on optimizing the locking mechanism of the widely used 2PL protocol in TXSQL. 
Although many efforts to optimize OLTP systems focus on different aspects, such as consensus protocols (e.g., \cite{DBLP:conf/sosp/ClementsKZMK13_Natto,DBLP:conf/osdi/MuNLL16_Janus,DBLP:conf/sigmod/YanYZLWSB18_Carousel}), data partitioning or disaggregation (e.g., \cite{DBLP:conf/edbt/QuamarKD13,taft2014store,DBLP:journals/pvldb/SerafiniTEPAS16_clay,DBLP:conf/osdi/Zhang0Y24_motor}), data scheduling (e.g., \cite{DBLP:journals/pvldb/CurinoZJM10,DBLP:conf/cloud/DasAA10-g-store,DBLP:journals/pvldb/DasNAA11-Albatross,DBLP:journals/pvldb/LuYM19,DBLP:conf/icde/ZhengZLYCPD24_Lion}), new hardware (e.g., \cite{DBLP:conf/sigmod/BoeschenB22_GaccO,DBLP:conf/osdi/QianG24_MPP_MVCC,DBLP:journals/vldb/ZhaoL0ZYZ0L0P24_RCBENCH}), and performance parameter tuning (e.g., \cite{DBLP:conf/sigmod/ZhangLZLXCXWCLR19,DBLP:journals/pvldb/LuCHB19,DBLP:journals/csur/HerodotouCL20}), they fall outside the scope of concurrency control in this paper and are orthogonal to our objectives.

\noindent \textbf{Hotspot optimizations.}
We have categorized hotspot optimizations into three types of scheduling.
(1) Thread-level scheduling~\cite{DBLP:journals/pvldb/RenTA12_vll}: This approach can complement our work by utilizing a thread pool to manage thread utilization, aiming to minimize thread contention. 
(2) Transaction-level scheduling \cite{DBLP:conf/vldb/KimuraGK12_early_lock_release,DBLP:conf/sigmod/GraefeLKTV13_CLV,DBLP:conf/sigmod/JungHFHY13,DBLP:journals/pvldb/YanC16_QURO,DBLP:conf/eurosys/ChenSJRLWZCC21_DAST}: This method incorporates lock queuing or minimizes lock holding time to mitigate lock contention. However, its effectiveness is limited in high-latency scenarios.
(3) Query-level scheduling (e.g., \cite{DBLP:conf/sosp/ZhangPZSAL13_chopping,DBLP:conf/sigmod/FaleiroTA14_lazy,DBLP:conf/sosp/XieSLAK015_Callas,DBLP:journals/pvldb/DingKG18,DBLP:journals/pvldb/LuYCM20_aria}): Deterministic protocols exemplify typical query-level scheduling methods. While these protocols can effectively eliminate lock contention, they necessitate prior knowledge of the read-write sets.
This paper has employed query-level scheduling for hotspot updates by automatically detecting hot data access and implemented a group locking mechanism, which groups hotspot updates and executes them serially without locking to enhance performance.

\section{Conclusion}
In this paper, we presented the Tencent Database System, TXSQL, and its optimizations towards high-contented workloads. We provide motivation, optimizations, and insights into TXSQL to address lock conflict issues. When dealing with hotspot data, we propose a group locking mechanism that groups hotspot data accesses and executes them serially without locking. Additionally, we describe its correctness in terms of deadlock prevention, rollback, and failure recovery. Through extensive analysis and performance evaluation, we demonstrate that TXSQL achieves performance improvements of up to 6.5x and 22.3x compared to state-of-the-art methods and systems, respectively. Our results also indicate the effectiveness and efficiency of TXSQL in real-world workloads.

\section{Acknowledgment}\label{sec:ack}
We would like to express our sincere gratitude to the reviewers for their invaluable comments and insightful suggestions.  We acknowledge the dedicated efforts of the TXSQL team, whose development work has been instrumental in the realization of this project.  

\balance


\bibliographystyle{ACM-Reference-Format}
\bibliography{reference}

\end{document}